\documentclass[lettersize,journal]{IEEEtran}
\usepackage{amsmath,amsfonts}
\usepackage{algorithmic}
\usepackage{algorithm}
\usepackage{array}
\usepackage{textcomp}
\usepackage{stfloats}
\usepackage{url}
\usepackage{verbatim}
\usepackage{graphicx}
\usepackage{cite}
\usepackage{enumerate}
\usepackage{amsmath}
\usepackage{graphicx}
\usepackage{booktabs}
\usepackage{subfigure}
\usepackage{epstopdf}
\usepackage{makecell}
\usepackage{threeparttable}
\usepackage[graphicx]{realboxes}
\usepackage[numbers,sort&compress]{natbib}
\usepackage[a4paper]{geometry}
\usepackage{float}
\begin{document}
\setlength{\abovedisplayskip}{4pt}
\setlength{\belowdisplayskip}{4pt}
\setlength{\abovedisplayshortskip}{4pt}
\setlength{\belowdisplayshortskip}{4pt}

\title{Nonparametric Stochastic Analysis of Dynamic Frequency in
Power Systems: A Generalized It\^o Process Model}

\author{Can Wan,~\IEEEmembership{Senior Member,~IEEE}, Yupeng Ren, and Ping Ju,~\IEEEmembership{Senior Member,~IEEE}
        % <-this % stops a space
\thanks{C. Wan, Y. Ren and P. Ju are with the College of Electrical Engineering, Zhejiang University, Hangzhou 310027, China (e-mail: canwan@zju.edu.cn, ryp22160310@zju.edu.cn,pju@zju.edu.cn).}% <-this % stops a space
}

\markboth{IEEE TRANSACTIONS ON POWER SYSTEMS}%
{Shell \MakeLowercase{\textit{et al.}}: A Sample Article Using IEEEtran.cls for IEEE Journals}


\maketitle

\begin{abstract}
The large-scale integration of intermittent renewable energy has brought serious challenges to the frequency security of power systems. In this paper, a novel nonparametric stochastic analysis method of system dynamic frequency is proposed to accurately analyze the impact of renewable energy uncertainty on power system frequency security, independent of any parametric distribution assumption. The nonparametric uncertainty of renewable generation disturbance is quantified based on probabilistic forecasting. Then, a novel generalized It\^o process is proposed as a linear combination of several Gaussian It\^o processes, which can represent any probability distribution. Furthermore, a stochastic model of power system frequency response is constructed by considering virtual synchronization control of wind power. On basis of generalized It\^o process, the complex nonlinear stochastic differential equation is transformed into a linear combination of several linear stochastic differential equations to approximate nonparametric probability distribution of the system dynamic frequency. Finally, the validity of the proposed method is verified by the single-machine system and IEEE 39-Bus system. 
\end{abstract}

\vspace{-0.3cm}

\section{Introduction}
\IEEEPARstart{W}{ith} the increasing penetration of renewable energy, renewable generation such as wind power and photovoltaic is gradually replacing the traditional synchronization units to become the main power source in modern power system \cite{ref1}. Unlike traditional synchronization units, renewable energy units do not have sufficient inertia and frequency response capabilities, of which the large-scale grid integration will inevitably result in inadequate inertia support and frequency adjustment capabilities of power system \cite{ref2}. In addition, the renewable energy generation has significant intermittence and fluctuation \cite{ref3}, which is regarded as an important factor that causes the dynamic frequency fluctuation of power system \cite{ref4}. The increase of power fluctuation randomness on both sides of generation and load further aggravates the potential frequency security problems of power system.

For frequency dynamic process analysis after system disturbance, simulation analysis is widely used in the offline analysis process \cite{ref5,ref6}, which has high accuracy but low computation speed. The analytic method presents the dynamic process of system frequency with mathematical analytic expression, which has fast computation speed and obvious advantages for frequency response characteristics \cite{ref7}. In \cite{ref8}, a low-order system frequency response (SFR) model is proposed to simplify the single-machine reheat thermal power unit model in terms of the corresponding time-domain analysis formula. An average system frequency model is established based on simplified assumptions in order to simplify and analyze the dynamic process of system frequency \cite{ref9}. A governor parameter aggregation method is proposed to equate the multi-machine system frequency response model to a single-machine model \cite{ref10}. Based on the open-loop SFR model, a frequency nadir calculation method is proposed by fitting the characteristics of the governor through the first-order inertia link \cite{ref11}, which uses a linear function to simulate the frequency deviation. In \cite{ref12}, a quadratic function is used to simulate the frequency deviation to realize the model open loop, and a more accurate frequency nadir calculation method is proposed accordingly via polynomial fitting of the governor characteristics.

The uncertainty of renewable generation is the main factor that leads to system frequency fluctuation in modern power systems. When studying the influence of renewable energy uncertainty on the system frequency, existing studies usually assume that the uncertainty of renewable energy generation obeys a parametric probability distribution such as Gaussian distribution \cite{ref13}, Weibull distribution \cite{ref14}, Beta distribution \cite{ref15}, etc. A SFR model under the stochastic disturbance of load and the stochastic error of measurement is proposed to estimate the intra-range probability of the system frequency \cite{ref16}, and a linear stochastic differential equation (SDE) is used to describe the above model. However, in \cite{ref16}, the stochastic variables are assumed as Gaussian white noises. A stochastic assessment function of automatic generation control is developed based on the series expansion method \cite{ref17}, which assumes that the stochastic resources satisfy the parametric distributions, such as Gaussian distribution and Beta distribution. However, the distribution of renewable generation uncertainty usually presents very severe polymorphism and fat tail characteristics, which is difficult to be accurately described by a specific parametric distribution \cite{ref3}. Nonparametric probabilistic forecasting can accurately quantify the uncertainty of renewable generation without any parametric distribution assumptions \cite{ref18}, such as quantiles \cite{ref3}. It becomes meaningful to analyze the stochastic characteristics of power system dynamic frequency considering nonparametric probabilistic forecasting of renewable generation.

This paper proposes a novel nonparametric stochastic analysis method to estimate the probability distribution of the system dynamic frequency under renewable power uncertainty without any parametric distribution assumptions. The nonparametric probabilistic forecasting based on quantiles is utilized to quantify the predictive uncertainty of renewable generation, which is further approximated by Gaussian mixture model (GMM). A generalized It\^o process model is proposed to describe any probability distributions with GMM decomposition. A unified SFR stochastic model considering the virtual synchronization control is constructed to describe the mapping relationship between the stochastic resources and system frequency. Based on the generalized It\^o process model, the complex nonlinear SDE of the proposed model can be transformed into a linear combination of linear SDEs to obtain a nonparametric probability distribution of system dynamic frequency. Finally, the effectiveness of the proposed method are illustrated by comprehensive case studies. In general, the main contributions of this paper are as follows

\begin{enumerate}
    \item 
    %1)
    A novel nonparametric stochastic analysis method based on generalized It\^o process (NSA-GIP) is proposed to avoid any distribution assumption for system dynamic frequency under renewable generation uncertainty.
    \item 
    %2)
    A generalized It\^o process model is proposed to describe any probability distributions based on the GMM decomposition.
    \item 
    %3)
    A unified SFR model is constructed to consider the effects of renewable generation on system dynamic frequency.
    \item 
    %4)
    An analytical calculation method is developed to convert the nonlinear SDE into a linear combination of linear SDEs based on the generalized It\^o process.
\end{enumerate}

The remainder of the paper is organized as follows. Section \ref{sec1} presents the generalized It\^o process of renewable generation. Section \ref{sec2} describes a nonparametric stochastic analysis method for system dynamic frequency. Comprehensive case studies are conducted in Section \ref{sec3} to verify the proposed method. Finally, Section \ref{sec4} concludes the paper.
\vspace{-0.3cm}
\section{Generalized It\^o Process of Renewable Generation}\label{sec1}
\subsection{Nonparametric Probabilistic Forecasting}
Renewable energy generation is regarded as one of the most important stochastic resources in modern power systems, of which the uncertainty can be accurately quantified by nonparametric probabilistic forecasting \cite{ref18}. Without loss of generality, renewable energy discussed in this paper mainly focuses on wind power. As an important form of nonparametric probabilistic forecasting, quantiles are used to describe the predictive uncertainty of renewable generation without any probability distribution assumption. The cumulative probability distribution function (CDF) of stochastic resources is defined as $F$, and the corresponding quantile $q^{\alpha}_{t}$ is defined as
\begin{equation}
\label{eq1}
{\rm{Pr}}(x_t\le q^{\alpha}_{t})=\alpha
\end{equation}
\begin{equation}
\label{eq2}
q_t^\alpha=F_t^{-1}(\alpha)
\end{equation}
where $\rm{Pr(\cdot)}$ represents the probability operator,$x_t$ is stochastic resource value at time $t$, $\alpha$ is the nominal proportion of the quantile $q^{\alpha}_{t}$. The series of predictive quantiles for stochastic resources can be obtained by direct quantile regression approach \cite{ref3}, expressed as
\begin{equation}
\label{eq3}
\hat{F}_t=\left\{\hat{q}_t^{\alpha_r} \mid 0 \leq \alpha_1<\cdots<\alpha_r<\cdots<\alpha_R \leq 1\right\}
\end{equation}
where $\hat{q}_t^{\alpha_r}$ represents the estimation of actual quantile $q^{\alpha}_{t}$, and $\hat{F}_t$ is the predictive quantile series with nominal proportion $\alpha_r$ need to be estimated.
\vspace{-0.3cm}

\subsection{Gaussian Mixture Model}
Gaussian mixture model (GMM) is a typical nonparametric model to describe probability distribution with the linear combination of several Gaussian distribution functions \cite{ref19}. Theoretically, GMM can fit any type of distribution. The probability density function (PDF) of the one-dimensional GMM $f_{\rm{GMM}}\left(x_t \mid \theta\right)$is expressed as
\begin{equation}
\label{eq4}
f_{\rm{GMM}}\left(x_t \mid \theta\right)=\sum_{i=1}^N \omega_i \frac{1}{\sqrt{2 \pi \sigma_i^2}} \exp \left[\frac{\left(x_t-\mu_i\right)^2} {2 \sigma_i^2}\right]
\end{equation}
where  $\omega_i$, $\mu_i$ and $\sigma_i^2$ represent the weight, expectation and variance of the \textit{i}-th sub-Gaussian component, respectively, and $N$ is the number of sub-Gaussian components in GMM. For each GMM, the set of parameters $\theta$ is unknown, expressed as:
\begin{equation}
\label{eq5}
\theta=\left\{\omega_i, \mu_i, \sigma_i^2\right\}_{i=1}^N
\end{equation}

Given the sub-Gaussian components number of GMM, the expected maximization (EM) algorithm is used to estimate probability model parameters with hidden variables \cite{ref20}, which consists of two steps in each iteration:
\begin{enumerate}[(1)]
\item Step 1 (E-step): Based on the current parameters, calculate the probability $\gamma_{i j}$ that each data $x_{t, j}$ comes from the \textit{i}-th Gaussian component, expressed as:
\begin{equation}
\label{eq6}
\gamma_{i j}^s=\frac{\omega_i^s \frac{1}{\sqrt{2 \pi\left(\sigma_i^2\right)^s}} \exp \left[\left(x_{t, j}-\mu_i^s\right)^2 / 2\left(\sigma_i^2\right)^s\right]}{\sum_{i=1}^N \omega_i^s \frac{1}{\sqrt{2 \pi\left(\sigma_i^2\right)^s}} \exp \left[\left(x_{t, j}-\mu_i^s\right)^2 / 2\left(\sigma_i^2\right)^s\right]}
\end{equation}
where $s$ is the number of iterations.
\item Step 2 (M-step): Assuming that the result of \eqref{eq6} is true, the estimated value of the parameter to be evaluated is calculated according to the maximum likelihood method. The formulas are given as follows:
\begin{equation}
\label{eq7}
\mu_i^{s+1}=\sum_{j=1}^M\left(\gamma_{i j}^s x_{t, j}\right) / \sum_{j=1}^M \gamma_{i j}^s
\end{equation}
\begin{equation}
\label{eq8}
\left(\sigma_i^2\right)^{s+1}=\sum_{j=1}^M \gamma_{i j}^s\left(x_{t, j}-\mu_i^{s+1}\right)^2 / \sum_{j=1}^M \gamma_{i j}^s
\end{equation}
\begin{equation}
\label{eq9}
\omega_i^{s+1}=\sum_{j=1}^M \gamma_{i j}^s / M
\end{equation}
\end{enumerate} 
where $M$ is the number of training sample $x_{t,j}$.

Repeat the calculation of the two steps until the result of the parameter to be solved converges, and the maximum likelihood solution of the GMM parameter is obtained.

The original EM algorithm relies on the selection of initial values, which has a slow iteration speed. In this paper, the initial dataset is divided into $N$ different classes $\Omega_N$ by the \textit{k}-means clustering algorithm, expressed as
\begin{equation}
\label{eq10}
\Omega_N=\left\{D^{(1)}, \cdots, D^{(i)}, \cdots, D^{(N)}\right\} i=1,2, \cdots, N
\end{equation}
\begin{equation}
\label{eq11}
D^{(i)}=\left\{x_{t, j}\right\}_{j=1}^{M_i}
\end{equation}

The expectation and variance of each class $D^{(i)}$ are used as the initial expectation and variance of the EM algorithm, and the data proportion $M_i/M$ in each class $D^{(i)}$ is used as the initial weight. This can reduce the sensitivity of the EM algorithm to initial values and the possibility of falling into local optima, while improving the iteration speed.

\vspace{-0.3cm}
\subsection{Generalized It\^o Process of Renewable Generation}
The classic It\^o process has been used to express wind power with an arbitrary parametric probability distribution \cite{ref17}, such as Gaussian, Beta, Weibull, etc. The SDE expression in It\^o process is consistent with the ordinary differential equation (ODE) of the SFR dynamic model, which would benefit for unifying description of stochastic SFR. In this paper, without loss of generality, wind power is considered as a stochastic resource, here let $P_{\rm{w}}$ represent the wind power, the following SDE can be obtained as
\begin{equation}
\label{eq12}
d P_{\mathrm{w}}=m\left(P_{\mathrm{w}}\right) d t+\tau\left(P_{\mathrm{w}}\right) d W_t
\end{equation}
where $m(\cdot)$ is the drift function driving $P_{\rm{w}}$ to a set point, $\tau(\cdot)$ is the diffusion function describing the stochastic characteristics, and $W_t$ is a standard Wiener stochastic process \cite{ref21}. It can be found that the It\^o process is actually an integral with respect to the standard Wiener stochastic process.

Given a certain parametric probability distribution, the corresponding It\^o process can be constructed via the method in \cite{ref21}. As the functions $m(\cdot)$ and $\tau(\cdot)$ are not unique, let the drift function $m(\cdot)$ be a linear function of the stochastic resource for easy calculation, expressed as
\begin{equation}
\label{eq13}
m\left(P_{\mathrm{w}}\right)=-\lambda_{\mathrm{w}} P_{\mathrm{w}}+c
\end{equation}
where $\lambda_{\mathrm{w}}$ is an optional positive real number, let it equal to $1$ in this paper, and $c$ is a constant related to parameters of the corresponding probability distribution, especially for Gaussian distribution $c$ is the corresponding expectation.

Then the diffusion function can be calculated by using the following formula, expressed as
\begin{equation}
\label{eq14}
\tau^2\left(P_{\mathrm{w}}\right)=2 \frac{\int_{-\infty}^{P_{\mathrm{w}}} m(z) p(z) d z}{p\left(P_{\mathrm{w}}\right)}
\end{equation}
where $p_(\cdot)$ is a given probability density function (PDF), and $z$ is an auxiliary variable \cite{ref21}.

The above classic It\^o process models can only describe specific parametric probability distributions by constructing drift and diffusion function of corresponding probability density functions, while the uncertainty of actual wind power cannot be accurately approximated by a specific parametric distribution. Therefore, GMM $f_{\mathrm{w}}\left(P_{\mathrm{w}} \mid \theta_{\mathrm{w}}\right)$ is utilized to describe the nonparametric probability distribution of wind power $P_{\rm{w}}$, which can be represented as a linear combination of several Gaussian distributions, expressed as
\begin{equation}
\label{eq15}
f_{\mathrm{w}}\left(P_{\mathrm{w}}\! \mid \!\theta_{\mathrm{w}}\right)\!=\!\sum_{i=1}^{n_{\mathrm{w}}} \omega_{\mathrm{w}, i} \frac{1}{\sqrt{2 \pi \sigma_{\mathrm{w}, i}^2}} \exp \left[\frac{\left(P_{\mathrm{w}}-\mu_{\mathrm{w}, i}\right)^2}{2 \sigma_{\mathrm{w}, i}^2}\right]
\end{equation}
\begin{equation}
\label{eq16}
\theta_{\mathrm{w}}=\left\{\omega_{\mathrm{w}, i}, \mu_{\mathrm{w}, i}, \sigma_{\mathrm{w}, i}^2\right\}_{i=1}^{n_{\mathrm{w}}}
\end{equation}
where $n_{\mathrm{w}}$ is the number of sub-Gaussian components, $\theta_{\mathrm{w}}$ represents the corresponding GMM parameter set, and the \textit{i}-th sub-Gaussian component is expressed as
\begin{equation}
\label{eq17}
f_i\left(P_{\mathrm{w}}\right)=\frac{1}{\sqrt{2 \pi \sigma_{\mathrm{w}, i}^2}} \exp \left[\frac{\left(P_{\mathrm{w}}-\mu_{\mathrm{w}, i}\right)^2} {2 \sigma_{\mathrm{w}, i}^2}\right]
\end{equation}
where $\mu_{\mathrm{w}, i}$ is the corresponding expectation, and $\sigma_{\mathrm{w}, i}^2$ is the corresponding variance. 

According to (12)-(14), It\^o process model corresponding to sub-Gaussian component of wind power $P_{\rm{w}}$ can be obtained, which are described as follow
\begin{equation}
\label{eq18}
d P_{\mathrm{w}}^{(i)}=\left(-P_{\mathrm{w}}+\mu_{\mathrm{w}, i}\right) d t+\sqrt{2 \sigma_{\mathrm{w}, i}^2} d W_t
\end{equation}

By decomposing the probability distribution of wind power $P_{\rm{w}}$ into $n_{\mathrm{w}}$ sub-Gaussian components, a generalized It\^o process, which can describe any probability distribution unlike classic It\^o processes, can be represented by a linear combination of $n_{\mathrm{w}}$ sub-Gaussian It\^o processes, expressed as
\begin{equation}
\label{eq19}
\left\{\begin{array}{c}
d P_{\mathrm{w}}^{(1)}=\left(-P_{\mathrm{w}}+\mu_{\mathrm{w}, 1}\right) d t+\sqrt{2 \sigma_{\mathrm{w}, 1}^2} d W_t \\
d P_{\mathrm{w}}^{(2)}=\left(-P_{\mathrm{w}}+\mu_{\mathrm{w}, 2}\right) d t+\sqrt{2 \sigma_{\mathrm{w}, 2}^2} d W_t \\
\vdots \\
d P_{\mathrm{w}}^{(i)}=\left(-P_{\mathrm{w}}+\mu_{\mathrm{w}, i}\right) d t+\sqrt{2 \sigma_{\mathrm{w}, i}^2} d W_t \\
\vdots \\
d P_{\mathrm{w}}^{\left(n_{\mathrm{w}}\right)}=\left(-P_{\mathrm{w}}+\mu_{\mathrm{w}, n_{\mathrm{w}}}\right) d t+\sqrt{2 \sigma_{\mathrm{w}, n_{\mathrm{w}}}^2} d W_t
\end{array}\right.
\end{equation}
where each sub-Gaussian It\^o process has a corresponding sub-Gaussian distribution, the weight of the \textit{i}-th sub-Gaussian It\^o process is the same as the \textit{i}-th sub-Gaussian component of wind power $P_{\rm{w}}$, which is equal to $\omega_{\mathrm{w}, i}$.

\vspace{-0.2cm}
\section{Nonparametric Stochastic Analysis of System Dynamic Frequency}\label{sec2}
\subsection{SFR Model With Wind Power Active Support}
The frequency response model of power systems, which mainly includes generator, turbine, governor and load, is a closed-loop control system. In this paper, it aims to study the overall system frequency response characteristics, so the dispersion of frequency and angle stability are not considered \cite{ref10}. The secondary frequency regulation and detailed nonlinear links such as complex governor control, amplitude limit and dead band are also neglected \cite{ref9}. The SFR model is widely used in system dynamic frequency analysis, where a complex power system dynamic model is simplified to a low-order system model, and all synchronous generators can be simplified to the closed-loop control model \cite{ref22}.
\begin{figure}[htbp]
\vspace{-0.4cm}
\centering
\includegraphics[width=2.5in]{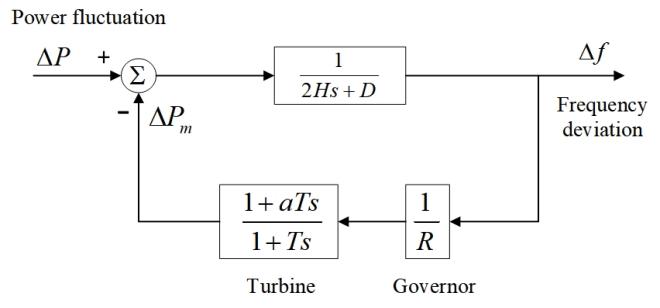}
\vspace{-0.3cm}
\caption{The typical SFR model.}
\label{fig1}
\vspace{-0.3cm}
\end{figure}

The transfer function of SFR model is shown in Fig. \ref{fig1}, where $H$ is the inertia time constant of synchronous generators, $D$ is the damping coefficient,  $a$ is the turbine characteristic coefficient, $T$ is the  turbine time constant, $R$ is the governor regulation coefficient, and $\Delta P_m$ is the mechanical power increment.

For the power system with high penetration of renewables, it is necessary to consider the frequency control of renewable generation. As an important way for renewable energy to participate in frequency regulation, virtual synchronous generator (VSG) can design control algorithm of grid-connected converter by simulating the external characteristics of the synchronous generator, such as inertia, damping and active power frequency regulation \cite{ref23}.

The inertial support power of traditional synchronous generators can be expressed as
\begin{equation}
\label{eq20}
\Delta P_e=-2 H \cdot \frac{d f}{d t} \cdot \frac{P_N}{f_0}
\end{equation}
where $\Delta P_e$ is the inertia support power of the synchronous generator, $P_N$ is the rated power of the synchronous generator and $f_0$ is the reference frequency of the power system.

With applying the virtual inertia control to the inverter \cite{ref23}, wind turbine VSG realizes inertial support by (\ref{eq20}), and its corresponding expression is represented as follow
\begin{equation}
\label{eq21}
\Delta P_{H, \mathrm{w}}=-2 H_{\mathrm{w}} \cdot \frac{d f}{d t} \cdot \frac{P_{\mathrm{w}}^{\max }}{f_0}
\end{equation}
where $\Delta P_{H, \mathrm{w}}$ is the inertia support power of the wind turbine VSG, $H_{\mathrm{w}}$ is the equivalent virtual inertia time constant of wind turbine VSG and $P_{\mathrm{w}}^{\max}$ is the rated power of the wind turbine VSG.

The wind turbine VSG can participate in primary frequency regulation by reserving part of spare power. The expression of primary frequency regulation power can be simplified as a linear function of the system dynamic frequency, expressed as
\begin{equation}
\label{eq22}
\Delta P_{k, \mathrm{w}}=-\frac{1}{\delta_{\mathrm{w}}} \cdot \frac{P_{\mathrm{w}}^{\max }}{f_0} \cdot \Delta f
\end{equation}
where $\Delta P_{k, \mathrm{w}}$ is the support power of the primary frequency regulation, $\delta_{\mathrm{w}}$ is the equivalent regulation coefficient of the wind turbine VSG, and $\Delta f$ is the system dynamic frequency denoting the system frequency deviation between the system frequency $f$ and its reference value $f_0$.

\begin{figure}[htbp]
\centerline{\includegraphics[scale=0.5]{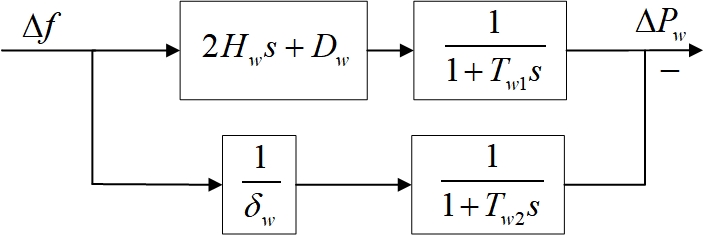}}
\caption{Active frequency support transfer function for Wind power VSG}
\label{figwindlink}
\end{figure}

To suppress the influence of unbalanced torque on the turbine, a first-order inertia link with a time constant of $T_{w}$ is usually added after \eqref{eq21} and \eqref{eq22} \cite{ref24}, which is shown in Fig. \ref{figwindlink}.

Due to the time constants of VSGs being approximately 2-3 orders of magnitude lower than that of synchronous generators, it can be assumed that $T_{w}\approx0$ \cite{ref25}. Therefore, a new VSG-SFR model can be proposed by considering the participation of wind turbine VSG in frequency control, shown in Fig. 2.
\begin{figure}[htbp]
\vspace{-0.2cm}
\centering
\includegraphics[scale=0.8]{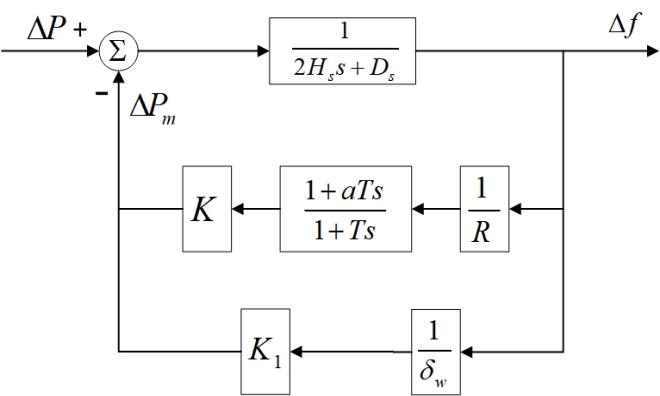}
\vspace{-0.3cm}
\caption{The VSG-SFR model.}
\label{fig2}
\vspace{-0.3cm}
\end{figure}

In Fig. \ref{fig2}, $H_s$ is the system equivalent inertia time constant, expressed by
\begin{equation}
\label{eq23}
H_s=K H+K_1 H_{\mathrm{w}}
\end{equation}
where $K$ is the proportion of synchronous generator capacity to the total capacity, and $K_1$ is the proportion of wind turbine capacity with VSGs.

\begin{figure}[htbp]
\vspace{-0.2cm}
\centering
\includegraphics[scale=0.8]{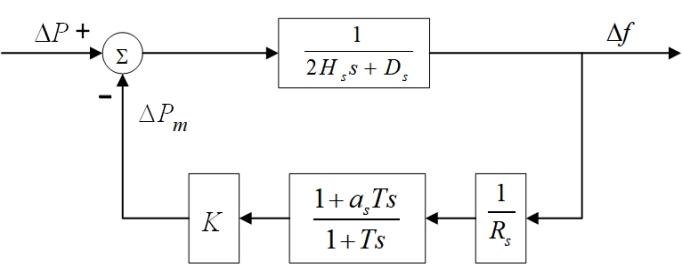}
\vspace{-0.1cm}
\caption{The simplified VSG-SFR model.}
\label{fig3}
\vspace{-0.4cm}
\end{figure}

The penetration of renewable energy can be obtained and expressed as
\begin{equation}
\label{eq24}
1-K=K_1+K_2
\end{equation}
where $K_2$ represents the proportion of wind turbine capacity without VSGs. After the transfer function operation, the VSG-SFR model can be transformed into the simplified model, shown in Fig. \ref{fig3}.

In the simplified model, the equivalent turbine characteristic coefficient $a_s$ and the equivalent governor regulation coefficient $R_s$ can be calculated by the formulas expressed as follow
\begin{equation}
\label{eq25}
\left\{\begin{array}{l}
a_s=\dfrac{K a+R K_1 / \delta_{\mathrm{w}}}{K+R K_1 / \delta_{\mathrm{w}}} \\
R_s=\dfrac{K}{K+R K_1 / \delta_{\mathrm{w}}} R
\end{array}\right.
\end{equation}

The simplified VSG-SFR model in Fig. \ref{fig3} can be expressed as an ordinary differential equation, given by
\begin{equation}
\label{eq26}
\setlength{\arraycolsep}{0.9pt}
\left[\begin{array}{c}
\dot{t_g} \\
\dot{\Delta f}
\end{array}\right]\!=\!\left[\begin{array}{cc}
-\dfrac{1}{T} & \dfrac{1-a_s}{R_s T} \\
-\dfrac{K}{2 H_s} & -\dfrac{D_s+K a_s / R_s}{2 H_s}
\end{array}\right]\!\left[\begin{array}{c}
t_g \\
\Delta f
\end{array}\right]\!+\!\left[\begin{array}{c}
0 \\
\dfrac{\Delta P}{2H_s}
\end{array}\right]
\end{equation}
where $t_g$ represents the system state of the governor.
\vspace{-0.3cm}
\subsection{Unified Stochastic SFR Model}
The system inertia model under stochastic resources can be formulated as
\begin{equation}
\label{eq27}
2 H_s \Delta \dot{f}=-\Delta P_t^m-D \Delta f+\Delta P
\end{equation}
where $\Delta P$ denotes the stochastic power fluctuation, expressed as
\begin{equation}
\label{eq28}
\Delta P=P_{\mathrm{w}}+P_{\mathrm{G}}-P_{\mathrm{L}}
\end{equation}
where $P_{\mathrm{G}}$ denotes the power of synchronous generators, $P_{\mathrm{w}}$ is the wind power which is a stochastic resource, and $P_{\mathrm{L}}$ represents the electricity load.

According to (\ref{eq12}), (\ref{eq27}) and (\ref{eq28}), a unified It\^o process of the system inertia model can be established and expressed as
\begin{equation}
\label{eq29}
\setlength{\arraycolsep}{0.9pt}
d\left[\begin{array}{l}
\Delta f \\
P_{\mathrm{w}}
\end{array}\right]\!=\!\left[\begin{array}{c}
\dfrac{-\Delta P_t^m-D \Delta f+\Delta P}{2 H_s} \\
m\left(P_{\mathrm{w}}\right)
\end{array}\right]\! d t\!+\!\left[\begin{array}{c}
0 \\
\tau\left(P_{\mathrm{w}}\right)
\end{array}\right]\! d W_t
\end{equation}

Considering the simplified governor model with wind turbine support and linearizing the drift function, a stochastic model of VSG-SFR can be obtained and shown in Fig. \ref{fig4}. According to (\ref{eq26}) and (\ref{eq28}), the stochastic model of VSG-SFR can be expressed as (30).
\begin{figure*}[!t]
\setcounter{equation}{29}
\begin{equation}
\label{eq30}
\setlength{\arraycolsep}{0.9pt}
d\left[\begin{array}{c}
t_g \\
\Delta f \\
P_{\mathrm{w}}
\end{array}\right]=\left(\left[\begin{array}{ccc}
-\dfrac{1}{T} & \dfrac{1-a_s}{R_s T} & 0 \\
-\dfrac{K}{2 H_s} & -\dfrac{D_s+K a_s / R_s}{2 H_s} & \dfrac{1}{2 H_s} \\
0 & 0 & -\lambda_w
\end{array}\right]\left[\begin{array}{c}
t_g \\
\Delta f \\
P_{\mathrm{w}}
\end{array}\right]+\left[\begin{array}{c}
0 \\
\dfrac{P_G-P_L}{2 H_s} \\
a_{\mathrm{w}}
\end{array}\right]\right) d t+\left[\begin{array}{c}
0 \\
0 \\
\tau\left(P_{\mathrm{w}}\right)
\end{array}\right] d W_t 
\end{equation}
\hrulefill
\vspace*{-0.2cm}
\end{figure*}
\begin{figure}[htbp]
\vspace{-0.1cm}
\centering
\includegraphics[scale=0.9]{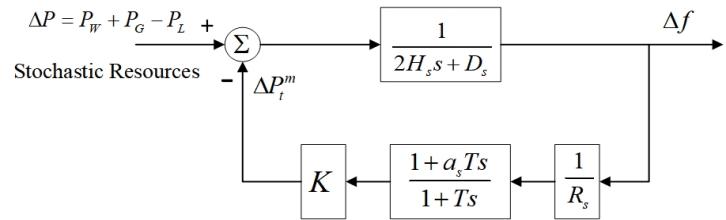}
\vspace{-0.2cm}
\caption{The VSG-SFR stochastic model.}
\label{fig4}
\vspace{-0.4cm}
\end{figure}

These abovementioned variables in (\ref{eq30}) are denoted as a vector $\textit{\textbf{X}}_t=\left[t_g, \Delta f, P_{\mathrm{w}}\right]^{\mathrm{T}}$, where the superscript “T” represents the transpose operation. The SDEs (\ref{eq29}) can be rewritten in a general form, given as
\begin{equation}
\label{eq31}
\left\{\begin{array}{l}
d \textit{\textbf{X}}_t=\left(\textit{\textbf{A}} \textit{\textbf{X}}_t+\textit{\textbf{c}}\right) d t+\tau\left(\textit{\textbf{X}}_t\right) d W_t \\
\textit{\textbf{X}}_0=\textit{\textbf{x}}_0
\end{array}\right.
\end{equation}
where $\textit{\textbf{x}}_0$ is the initial value of $\textit{\textbf{X}}_t$. The state matrix can be expressed by
\begin{equation}
\label{eq32}
\textit{\textbf{A}}=\left[\begin{array}{ccc}
-\dfrac{1}{T} & \dfrac{1-a_s}{R_s T} & 0 \\
-\dfrac{K}{2 H_s} & -\dfrac{D_s+K a_s / R_s}{2 H_s} & \dfrac{1}{2 H_s} \\
0 & 0 & -\lambda_{\mathrm{w}}
\end{array}\right]
\end{equation}

The diffusion function vector can be expressed by
\begin{equation}
\label{eq33}
\tau\left(\textit{\textbf{X}}_t\right)=\left[\begin{array}{lll}
0 & 0 & \tau\left(P_{\mathrm{w}}\right)
\end{array}\right]^{\mathrm{T}}
\end{equation}
\vspace{-0.3cm}
\subsection{Nonlinear SDE Solution Based on Generalized It\^o process}
The complex stochastic characteristic of the wind power $P_{\mathrm{w}}$ can be well expressed in the form of a nonparametric distribution. For nonparametric probability distributions, the diffusion function $\tau\left(\textit{\textbf{X}}_t\right)$ of SDE \eqref{eq19} is an unknown nonlinear function which cannot be solved by neither direct solution method \cite{ref16} nor series expansion method \cite{ref17}. Based on the generalized It\^o process model \eqref{eq11}, the original complex nonlinear SDE \eqref{eq19} can be converted into a linear combination of $n_{\mathrm{w}}$ linear SDEs corresponding to $n_{\mathrm{w}}$ Gaussian distributions \cite{ref26}. The \textit{i}-th SDE is expressed as (34).
\begin{figure*}[!t]
\setcounter{equation}{33}
\begin{equation}
\label{eq34}
\setlength{\arraycolsep}{1pt}
d\left[\begin{array}{c}
t_g \\
\Delta f \\
P_{\mathrm{w}}
\end{array}\right]^{(i)}\!=\!\left(\left[\begin{array}{ccc}
-\dfrac{1}{T} & \dfrac{1-a_s}{R_s T} & 0 \\
-\dfrac{K}{2 H_s} & -\dfrac{D_s+K a_s / R_s}{2 H_s} & \dfrac{1}{2 H_s} \\
0 & 0 & -1
\end{array}\right]\!\left[\begin{array}{c}
t_g \\
\Delta f \\
P_{\mathrm{w}}
\end{array}\right]\!+\!\left[\begin{array}{c}
0 \\
\dfrac{P_{\mathrm{G}}-P_{\mathrm{L}}}{2 H_s} \\
\mu_{\mathrm{w}, i}
\end{array}\right]\right)\!d t\!+\!\left[\begin{array}{c}
0 \\
0 \\
\sqrt{2 \sigma_{\mathrm{w}, i}^2}
\end{array}\right]\! d W_t \quad i=1,2, \cdots, n_{\mathrm{w}}
\end{equation}
\hrulefill
\vspace*{-0.2cm}
\end{figure*}

Write \eqref{eq34} in a composite form as follow
\begin{equation}
\label{eq35}
\left\{\begin{array}{l}
d \textit{\textbf{X}}_t^i=\left(\textit{\textbf{A}} \textit{\textbf{X}}_t^i+\textit{\textbf{c}}_i\right) d t+\textit{\textbf{B}}_i d W_t \\
\textit{\textbf{X}}_0=\textit{\textbf{x}}_0
\end{array}\right.
\end{equation}
where $\textit{\textbf{B}}_i$ and $\textit{\textbf{c}}_i$ are constant vectors of the \textit{i}-th SDE, and $\textit{\textbf{X}}_t^i$ is the \textit{i}-th component of $\textit{\textbf{X}}_t$.

According to the linearized stochastic theory in \cite{ref26}, rewrite the composite formula \eqref{eq35} as
\begin{equation}
\label{eq36}
e^{-\textit{\textbf{A}} t} d \textit{\textbf{X}}_t^i-e^{-\textit{\textbf{A}} t}\left(\textit{\textbf{A}} \textit{\textbf{X}}_t^i+\textit{\textbf{c}}_i\right) d t=e^{-\textit{\textbf{A}} t} \textit{\textbf{B}}_i d W_t
\end{equation}

According to the It\^o formula, differentiate $e^{-\textit{\textbf{A}} t} d \textit{\textbf{X}}_t^i$, it can be obtained as follow
\begin{equation}
\label{eq37}
d\left(e^{-\textit{\textbf{A}} t} \textit{\textbf{X}}_t^i\right)=e^{-\textit{\textbf{A} t}}(-\textit{\textbf{A}}) \textit{\textbf{X}}_t^i d t+e^{-\textit{\textbf{A}} t} d \textit{\textbf{X}}_t^i
\end{equation}

Substituting \eqref{eq35} into \eqref{eq36}, it can be obtained as 
\begin{equation}
\label{eq38}
d\left(e^{-\textit{\textbf{A}} t} \textit{\textbf{X}}_t^i\right)=e^{-\textit{\textbf{A}} t} \textit{\textbf{c}}_i d t+e^{-\textit{\textbf{A}} t} \textit{\textbf{B}}_i d W_t
\end{equation}

By integrating both sides of \eqref{eq38}, the solution of \eqref{eq34} can be deduced as
\begin{equation}
\label{eq39}
\textit{\textbf{X}}_t^i=e^{\textit{\textbf{A}} t}\left(\textit{\textbf{X}}_0+\textit{\textbf{A}}^{-1}\textit{\textbf{c}}_i\right)-\textit{\textbf{A}}^{-1}\textit{\textbf{c}}_i+\int_0^t e^{\textit{\textbf{A}} (t-s)} \textit{\textbf{B}}_i d W_s
\end{equation}
where $e^{\textit{\textbf{A}} t}$ is an exponential function of the \textit{n}th-order square matrix $\textit{\textbf{A}} t$.

The system states $\textit{\textbf{X}}_t^i$ can be proved to follow the Gaussian distribution, because $ \int_0^t e^{\textit{\textbf{A}} (t-s)}  \textit{\textbf{B}}_i d W_s$ follows the Gaussian distribution \cite{ref27}.

The expectations and variances of \eqref{eq39} can be calculated by
\begin{equation}
\label{eq40}
\mathrm{E}\left\{\textit{\textbf{X}}_t^i\right\}=e^{\textit{\textbf{A}} t}\left(\textit{\textbf{X}}_0+\textit{\textbf{A}}^{-1}\textit{\textbf{c}}_i\right)-\textit{\textbf{A}}^{-1}\textit{\textbf{c}}_i
\end{equation}
\begin{equation}
\label{eq41}
\operatorname{var}\left\{\textit{\textbf{X}}_t^i\right\}=\textit{\textbf{P}}\left\{\left[\textit{\textbf{P}}^{-1} \textit{\textbf{B}} \textit{\textbf{B}}^{\mathrm{T}}\left(\textit{\textbf{P}}^{-1}\right)^{\mathrm{T}}\right] \circ \textit{\textbf{J}}\right\} \textit{\textbf{P}}^{\mathrm{T}}
\end{equation}
\begin{equation}
\label{eq42}
\textit{\textbf{J}}(k, j)=\left[e^{\left(\lambda_k+\lambda_j\right) t}-1\right] /\left(\lambda_k+\lambda_j\right)
\end{equation}
\begin{equation}
\label{43}
\textit{\textbf{P}} \Lambda \textit{\textbf{P}}^{-1}=\textit{\textbf{A}}
\end{equation}
where $\mathrm{E}\left\{\textit{\textbf{X}}_t^i\right\}$ is the expectation vector of $\textit{\textbf{X}}_t^i$, $\operatorname{var}\left\{\textit{\textbf{X}}_t^i\right\}$ is the variance matrix of $\textit{\textbf{X}}_t^i$, $\lambda_k$ and $\lambda_j$ are the \textit{k}-th and \textit{j}-th eigenvalue of $\textit{\textbf{A}}$, $\textit{\textbf{P}}$ is a square matrix whose columns are the independent eigenvectors of $\textit{\textbf{A}}$, $\Lambda$ is a square matrix whose \textit{k}-th or \textit{j}-th diagonal entries are the $\lambda_k$ or $\lambda_j$, and “$\circ$” is the Hadamard product. The above derivation can be referred to \cite{ref27}.

At time $t$, the \textit{i}-th sub-Gaussian component PDF of the system dynamic frequency $\Delta f_t$ can be described as
\begin{equation}
\label{eq44}
f_i\left(\Delta f_t\right)=\frac{1}{\sqrt{2 \pi \sigma_{\Delta f, i}^2}} \exp \left[\left(\Delta f_t-\mu_{\Delta f, i}\right)^2 / 2 \sigma_{\Delta f, i}^2\right]
\end{equation}
where $\mu_{\Delta f, i}$ is the expectation of the \textit{i}-th sub-Gaussian component of $\Delta f_t$, which is actually the second entry of $\mathrm{E}\left\{\textit{\textbf{X}}_t^i\right\}$, and $\sigma_{\Delta f, i}^2$ is the variance of the \textit{i}-th sub-Gaussian component of $\Delta f_t$, which is just the second diagonal entry of the $\operatorname{var}\left\{\textit{\textbf{X}}_t^i\right\}$.

The VSG-SFR model proposed in this paper is actually a linear and time-invariant system. The weight of the \textit{i}-th sub-Gaussian component of the system dynamic frequency $\Delta f_t$ remains the same as the weight of the \textit{i}-th sub-Gaussian component of wind power $P_{\mathrm{w}}$ according to the linear invariance of GMM \cite{ref28} and the stochastic dynamics theory \cite{ref29}. The PDF of system dynamic frequency component obtained from each SDE \eqref{eq44} is weighted and integrated to obtain the general probability distribution of the system dynamic frequency $\Delta f_t$, expressed as
\begin{equation}
\label{eq45}
\begin{split}
&f_{\mathrm{PDF}}\left(\Delta f_t\right)=\sum_{i=1}^{n_{\mathrm{w}}} \omega_{\mathrm{w},i} \frac{1}{\sqrt{2 \pi \sigma_{\Delta f, i}^2}} \exp \left[\frac{\left(\Delta f_t-\mu_{\Delta f, i}\right)^2}{2 \sigma_{\Delta f, i}^2}\right] \\
&i=1,2, \cdots, n_{\mathrm{w}}
\end{split}
\end{equation}

Accordingly, the CDF of the system dynamic frequency $\Delta f_t$ is given as follow
\begin{equation}
\label{eq46}
\begin{split}
& F_{\mathrm{CDF}}\left(\Delta f_t\right)\!=\\ & \!\sum_{i=1}^{n_{\mathrm{w}}}\left\{\omega_{\mathrm{w},i}\!\int_{-\infty}^{\Delta f_t}\! \frac{1}{\sqrt{2 \pi \sigma_{\Delta f, i}^2}}\! \exp \left[\frac{\left(\varphi-\mu_{\Delta f, i}\right)^2} {2 \sigma_{\Delta f, i}^2}\right]\! d \varphi\right\} \\
& i=1,2, \cdots, n_{\mathrm{w}}
\end{split}
\end{equation}

\vspace{-0.3cm}
\subsection{Implementation Procedure}
\begin{figure}[htbp]
\vspace{-0.2cm}
\centering
\includegraphics[scale=0.55]{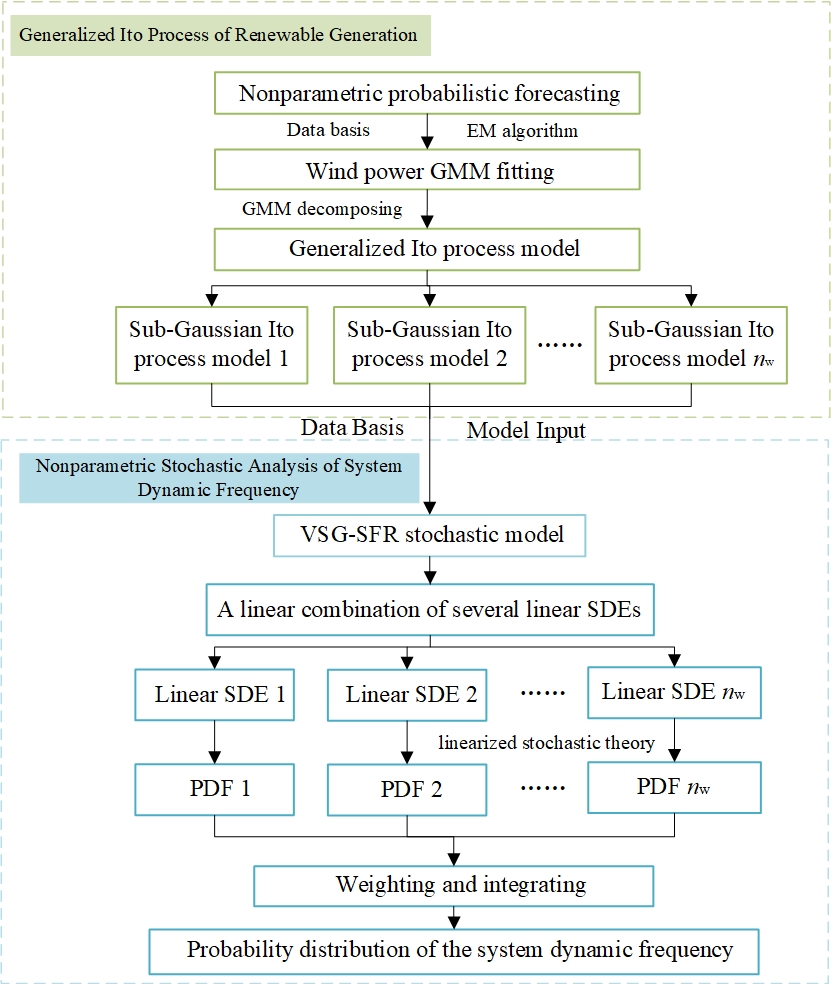}
\vspace{-0.2cm}
\caption{The procedure of the NSA-GIP method.}
\label{fig5}
\vspace{-0.3cm}
\end{figure}

The procedure of the proposed NSA-GIP method for power system dynamic frequency is shown in Fig. \ref{fig5}. in general, there are five steps given as follows
\begin{enumerate}[ Step 1)]
    \item 
    %1)
    Based on nonparametric probabilistic forecasting \cite{ref3}, the uncertainty of future wind power $P_{\rm{w}}$ can be quantified by quantiles without needs of any specific parametric distribution.
    \item 
    %2)
    According to the probabilistic forecasting results of Step 1, GMM of wind power $P_{\rm{w}}$ can be established, and the EM algorithm initialized with $k$-means clustering algorithm is used to obtain the parameter set $\theta_{\rm{w}}$.
    \item 
    %3)
    According to the GMM of wind power $P_{\rm{w}}$, the generalized It\^o process model is established as a linear combination of several Gaussian It\^o process models.
    \item 
    %4)
    A simplified VSG-SFR model is established and described via a nonlinear SDE, and the complex nonlinear SDE is decomposed into a linear combination of several linear SDEs based on the generalized It\^o process of Step 3.
    \item 
    %4)
    Each linear SDE is solved to obtain each sub-Gaussian component of the system dynamic frequency.
     \item 
    %4)
    The sub-Gaussian components of system dynamic frequency obtained from linear SDEs in Step 5 are weighted and integrated to obtain the general probability distribution of the system dynamic frequency.
\end{enumerate}

\section{Case Study}\label{sec3}

\subsection{Accuracy and Validity of VSG-SFR Model}
At first, the accuracy of the proposed VSG-SFR model needs to be verified by comparing with the uniform system frequency response (USFR) model proposed in \cite{ref25} based on Matlab/Simulink. Meanwhile, the improvement of the proposed model in suppressing frequency changes compared to SFR model also needs to be verified. The VSG-SFR model is simulated as the first case, which only has a thermal machine and a wind turbine. The parameters of the model are derived from the parameter aggregation of the USFR model and shown in Table \ref{tab1}.  
\begin{table}[htbp]
\vspace{-0.3cm}
\caption{Parameters of The Simplified VSG-SFR Model}
\label{tab1}
\centering
\vspace{-0.2cm}
\begin{tabular}{c c c c c c c c}
\toprule[1.1pt]
$1/R$ & $H$ & $a$ & $T$ & $D$ & $\delta_{\rm{w}}$ & $H_{\rm{w}}$ \\
\midrule
16.5 & 4.96 & 0.278 & 10 & 1.2 & 0.05 & 2 \\
\bottomrule[1.1pt]
\end{tabular}
\vspace{-0.1cm}
\end{table}

\begin{figure}[htbp]
\vspace{-0.3cm}
\centering
\subfigure[20\% wind power penetration]
{
    \begin{minipage}[b]{.4\linewidth}
        \flushleft
        \includegraphics[scale=0.13]{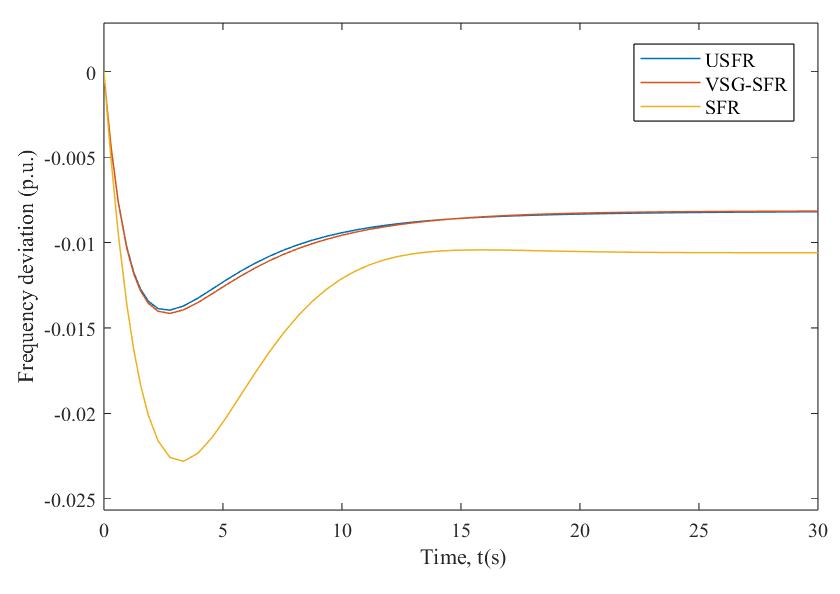}
    \end{minipage}
}
\subfigure[30\% wind power penetration]
{
    \begin{minipage}[b]{.4\linewidth}
        \flushleft
        \includegraphics[scale=0.13]{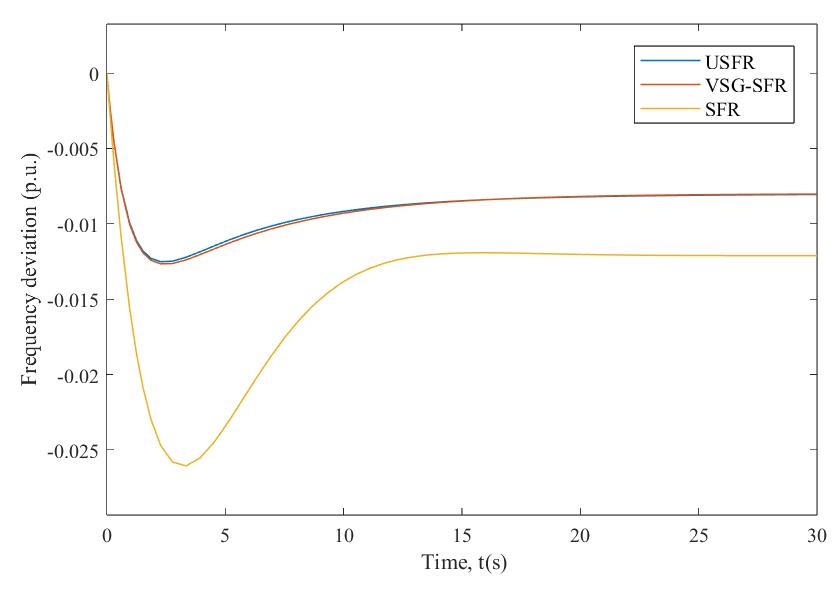}
    \end{minipage}
}
\subfigure[40\% wind power penetration]
{
    \begin{minipage}[b]{.4\linewidth}
        \flushleft
        \includegraphics[scale=0.13]{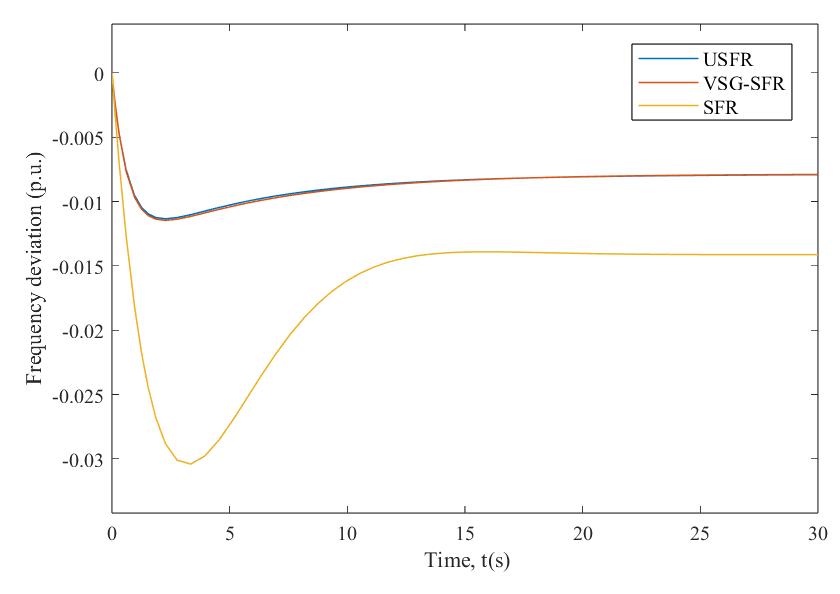}
    \end{minipage}
}
\subfigure[50\% wind power penetration]
{
    \begin{minipage}[b]{.4\linewidth}
        \flushleft
        \includegraphics[scale=0.13]{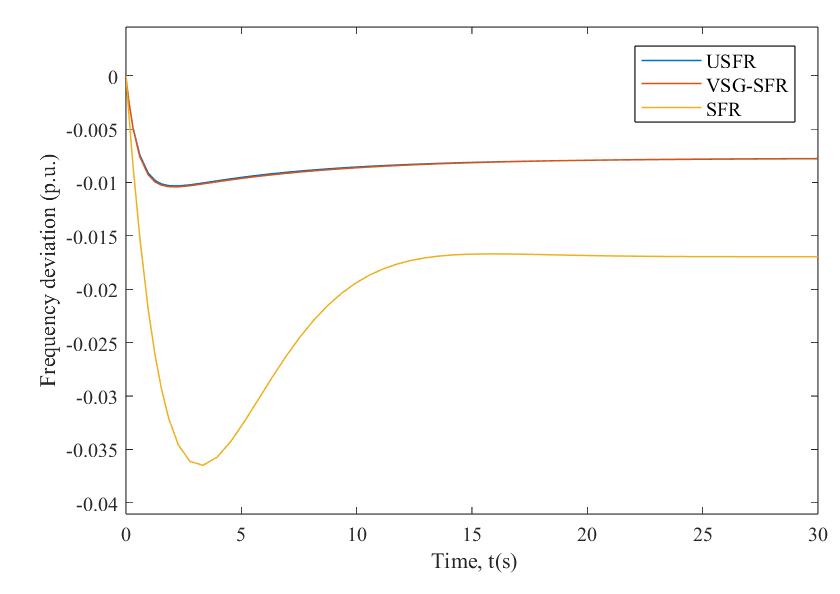}
    \end{minipage}
}
\vspace{-0.3cm}
\caption{The system frequency response characteristics under different renewable energy penetrations.}
\label{fig6}
\vspace{-0.2cm}
\end{figure}

The VSG-SFR model, USFR model and traditional SFR model without active support of wind power are respectively subject to a constant power disturbance, while changing the proportion of wind power to obtain the system frequency response curves under different renewable energy penetration levels, including 20\%, 30\%, 40\% and 50\%, which are shown in Fig. 6. It can be seen from Fig. \ref{fig6} that no matter how the penetration changes, the VSG-SFR model is always almost identical to the USFR Model, with maximum error ranging from 1.4\% to 3.2\%, which is sufficient to demonstrate the accuracy of the proposed model. Under the same renewable penetration level, the VSG-SFR model can increase the frequency nadir and quasi-steady frequency and reduce the rate of change of the frequency through the active support of wind power. By comparing the differences of Fig. \ref{fig6} (a)-(d), it can be found that with the increasing penetration of renewable energy, the VSG-SFR model has better regulation effect on the system dynamic frequency compared with the traditional SFR model. It indicates the validity of the VSG-SFR model for power system with high penetration of renewable energy.

\vspace{-0.3cm}
\subsection{Validity of NSA-GIP Method}
\begin{figure}[htbp]
\vspace{-0.1cm}
\centering
\includegraphics[scale=0.3]{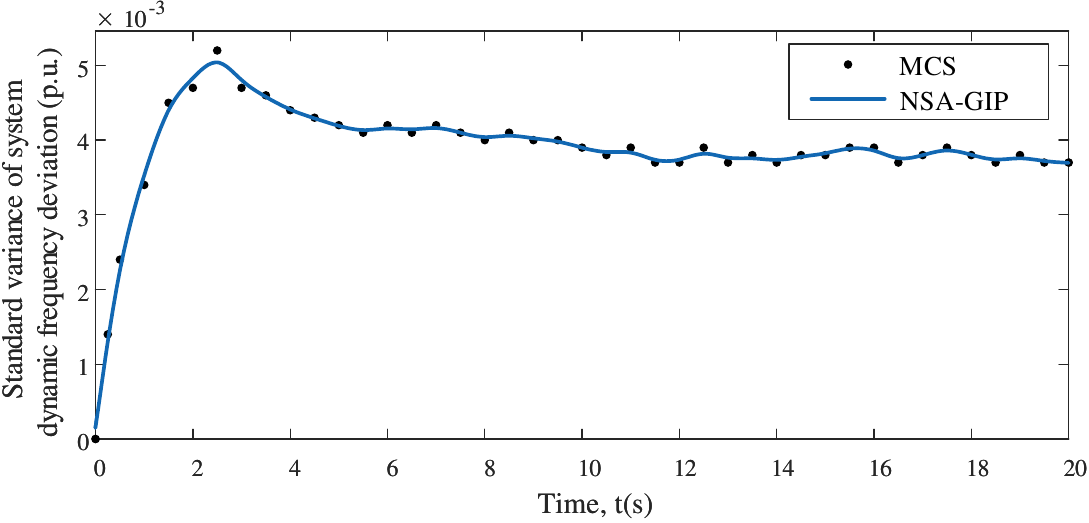}
\vspace{-0.3cm}
\caption{The standard variance of the system dynamic frequency deviation over time.}
\label{fig7}
\vspace{-0.2cm}
\end{figure}

To verify the validity of the proposed method in the VSG-SFR model, the standard variances of the system dynamic frequency deviation over time can be calculated based on Monte Carlo simulation (MCS) and the proposed method. The number of MCS simulations is set to 20000. The standard variance curve of the system dynamic frequency deviation is depicted in Fig. \ref{fig7}. It can be seen from Fig. \ref{fig7} that the results of the proposed method match well with those of MCS. Moreover, the standard deviation of frequency deviation is very small near the initial time, and its probability distribution is concentrated near the initial value. However, after a certain time, the standard deviation of frequency deviation will tend to be stable, and its probability distribution will converge to a stable distribution.

In order to verify the effectiveness and accuracy of the proposed method, it is necessary to compare whether the PDF and CDF calculated by the proposed method are consistent with those obtained by MCS. Here, the PDF and CDF of the system dynamic frequency deviation at 5s are selected for comparison with those obtained by MCS, which is shown in Fig. 8. 
\begin{figure}[htbp]

\centering
\subfigure[PDF]
{
    \begin{minipage}[b]{.4\linewidth}
        \flushleft
        \includegraphics[scale=0.138]{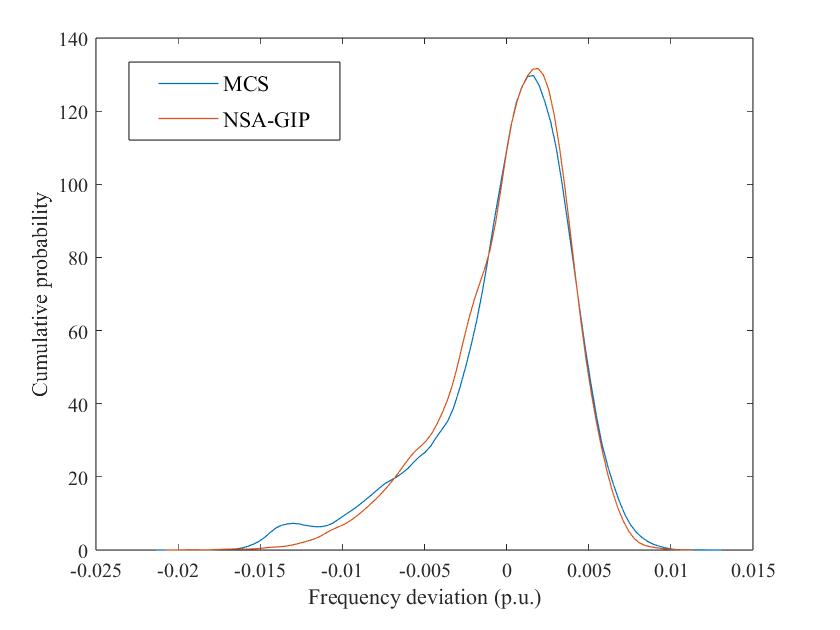}
    \end{minipage}
}
\subfigure[CDF]
{
    \begin{minipage}[b]{.4\linewidth}
        \flushleft
        \includegraphics[scale=0.251]{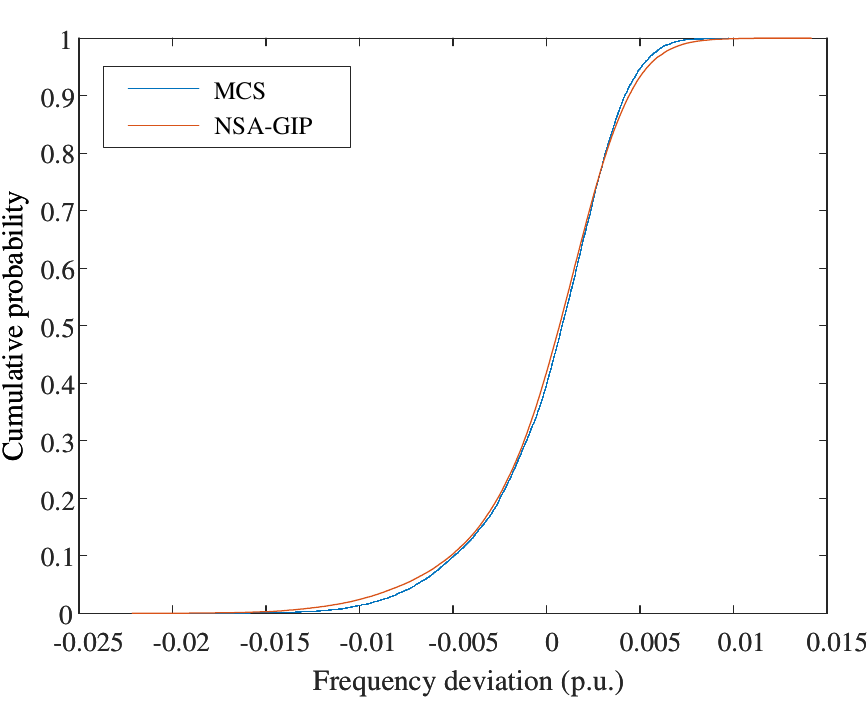}
    \end{minipage}
}
\vspace{-0.3cm}
\caption{Probability distribution of the system dynamic frequency deviation at 5s.}
\label{fig8}
\vspace{-0.2cm}
\end{figure}

The standard variances of MCS and NSA-GIP are respectively 0.0039 and 0.0040, whose error rate is only 2.56\%. Moreover, it can be seen that the PDF and CDF curves of the proposed method match very well with those of MCS.

\vspace{-0.3cm}
\subsection{Impacts of VSG-SFR Model Parameters}
The validity of the proposed method under different VSG-SFR model parameters needs to be further verified. Since the uncertainty of dynamic frequency with the nonparametric probability distribution, the proportion deviation (PD) \cite{ref30} is introduced herein to comprehensively measure the accuracy of the proposed method. The proportion deviation of the quantile $q_x^\alpha$ is defined as
\begin{equation}
\label{eq47}
P D_x^\alpha=\frac{1}{N} \sum_{i=1}^N \eta_i-\alpha
\end{equation}
where $\alpha$ represents the nominal proportion level, $N$ is the total number of samples, and $\eta_i$ is the indicator function of the \textit{i}-th sample, described as follow:
\begin{equation}
\label{eq48}
\eta_i=1\left(x_i \leq q_x^\alpha\right)
\end{equation}
where $x_i$ is the \textit{i}-th sample of the system dynamic frequency. Obviously, the closer the quantile deviation is to 0, the more accurate the estimated probability distribution is.

Fig. 9 shows the PD curves of the system dynamic frequency probability distribution under different VSG-SFR model parameters, including inertia time constant $H$, turbine characteristic coefficient $a$, governor regulation coefficient $R$, renewable energy penetration $1$-$K$, virtual inertia time constant $H_{\rm{w}}$, and wind turbine virtual regulation coefficient $\delta_{\rm{w}}$.
\begin{figure*}[htbp]
\vspace{-0.5cm}
\centering
\subfigure[$H$]
{
    \begin{minipage}[b]{.3\linewidth}
        \centering
        \includegraphics[scale=0.16]{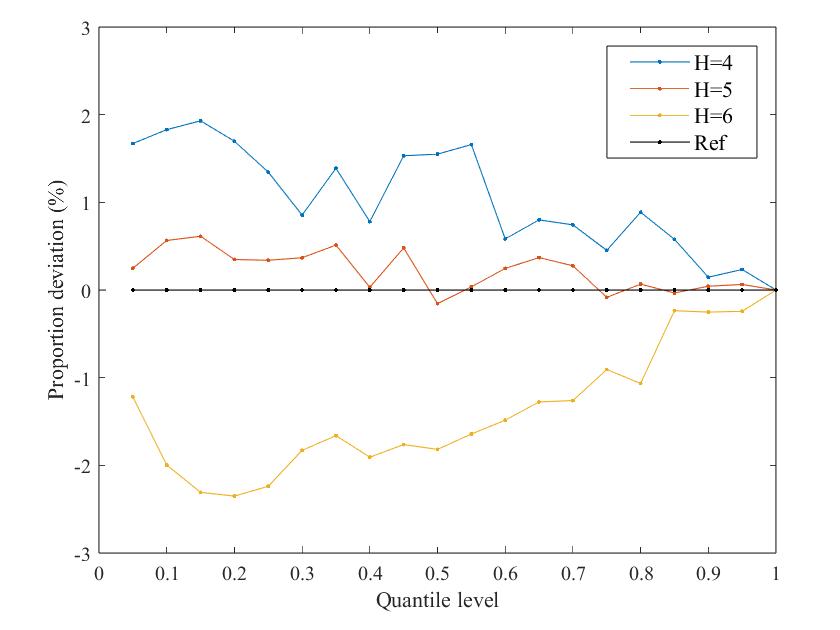}
    \end{minipage}
}
\subfigure[$a$]
{
    \begin{minipage}[b]{.3\linewidth}
        \centering
        \includegraphics[scale=0.16]{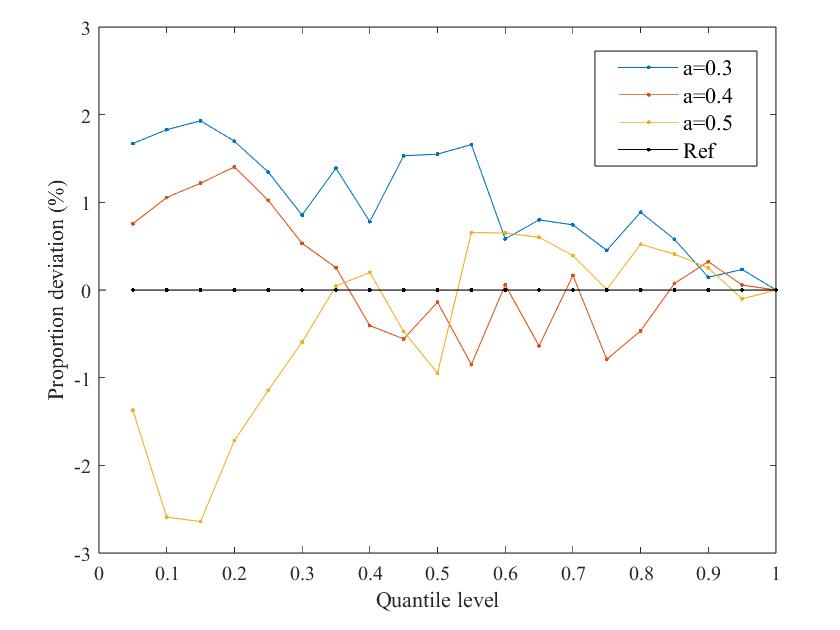}
    \end{minipage}
}
\subfigure[$R$]
{
    \begin{minipage}[b]{.3\linewidth}
        \centering
        \includegraphics[scale=0.16]{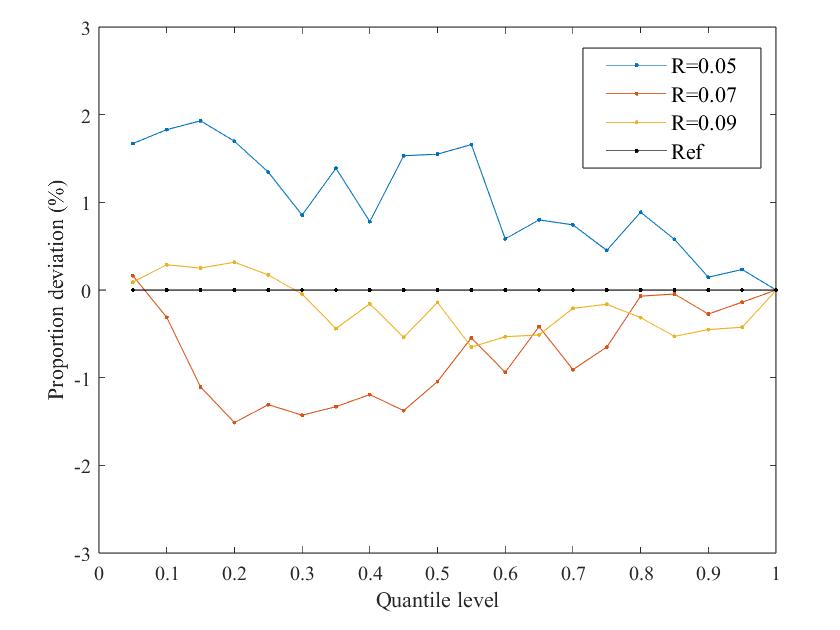}
    \end{minipage}
}
\subfigure[$K$]
{
    \begin{minipage}[b]{.3\linewidth}
        \centering
        \includegraphics[scale=0.16]{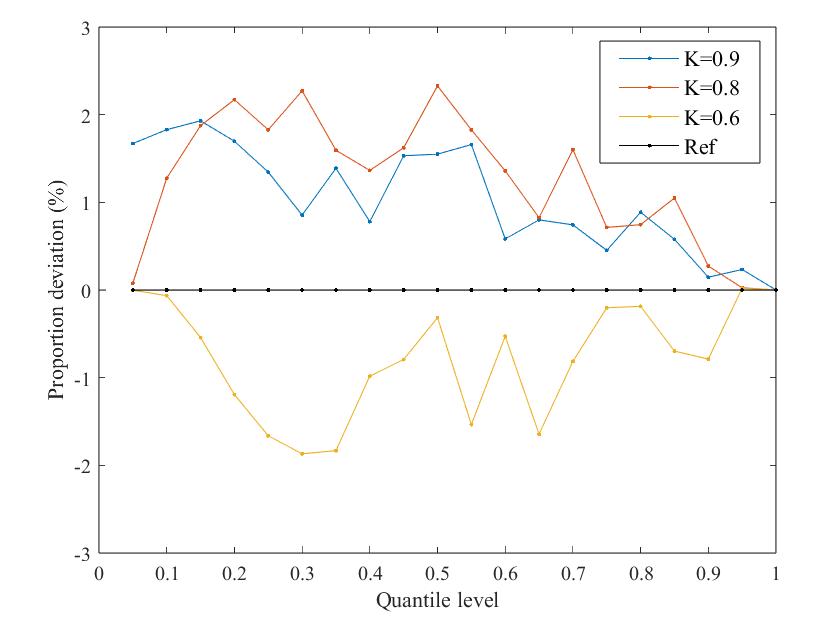}
    \end{minipage}
}
\subfigure[$H_{\rm{w}}$]
{
    \begin{minipage}[b]{.3\linewidth}
        \centering
        \includegraphics[scale=0.16]{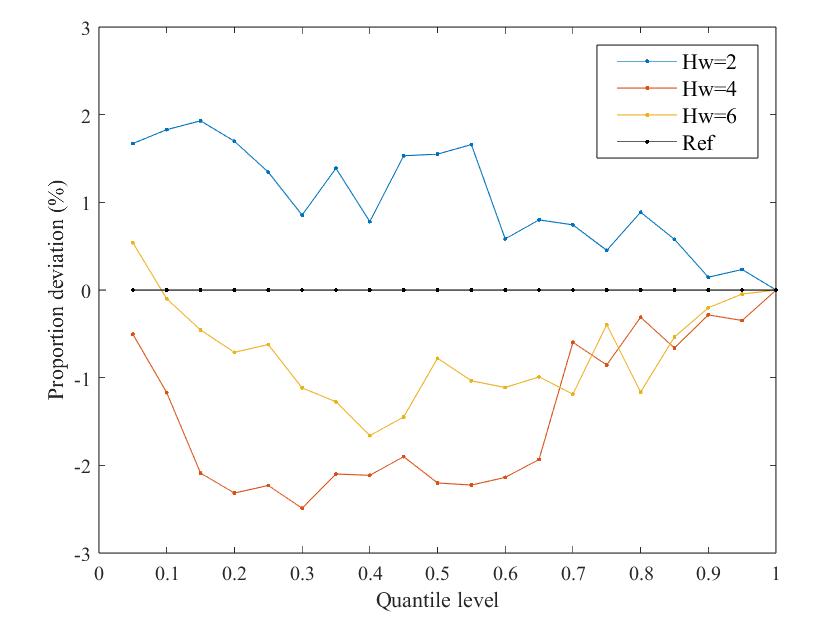}
    \end{minipage}
}
\subfigure[$\delta_{\rm{w}}$]
{
    \begin{minipage}[b]{.3\linewidth}
        \centering
        \includegraphics[scale=0.16]{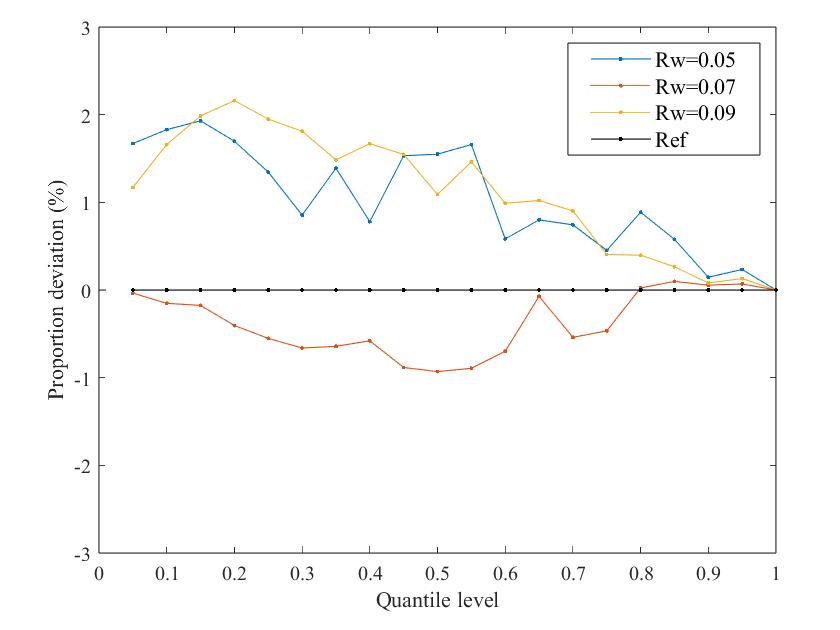}
    \end{minipage}
}
\vspace{-0.2cm}
\caption{Proportion deviation of the system dynamic frequency deviation under different model parameters.}
\label{fig9}
\vspace{-0.2cm}
\end{figure*}

It can be seen that the error between the dynamic frequency probability distribution obtained by the proposed method and the reference distribution is within 3\% regardless of the model parameters, so the effectiveness of the NSA-GIP method is almost unaffected by parameter changes.

\vspace{-0.1cm}
\subsection{Validity of NSA-GIP Method on Larger System}
\subsubsection{Case Settings}The larger case is conducted on the IEEE 39-Bus system, which includes 10 generators and 46 lines, as shown in Fig. 10. The parameters of this system can be found in \cite{ref31}. In this paper, the generators G1-G7 are thermal power stations, and the generators G8, G9 and G10 are wind farms, whose data comes from measured values in Denmark. The renewable energy penetration in this case is about 30\%. Since this paper only considers the overall system dynamic frequency characteristics, it is necessary to aggregate parameters of multi-machines in the system using the approach in \cite{ref11}.
\begin{figure}[htbp]
\vspace{-0.2cm}
\centering
\includegraphics[scale=0.4]{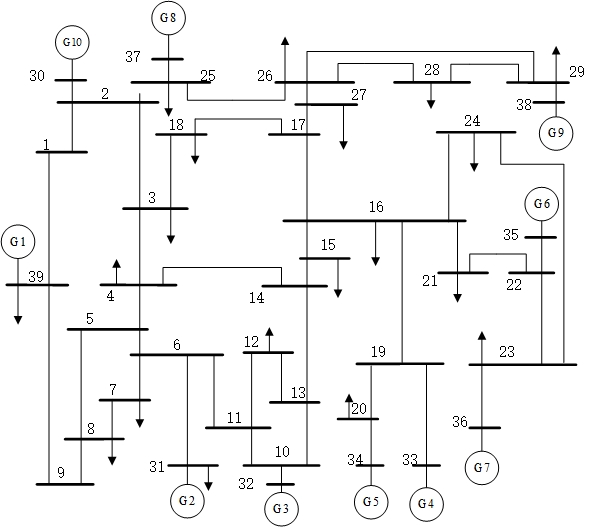}
\vspace{-0.3cm}
\caption{The IEEE 39-Bus system.}
\label{fig10}
\vspace{-0.2cm}
\end{figure}

\subsubsection{Simulation Results} To further verify the effectiveness and accuracy of the proposed method, the results obtained through MCS are considered as benchmarks. The simulation number of MCS here is set to 20000. The analytical methods including the direct solution method (DSM) \cite{ref16} and the series expansion method (SEM) \cite{ref17} are also implemented for comprehensive comparisons. The DSM assumes that the probability distribution of the input power fluctuation is approximately Gaussian. The SEM uses the series expansion to estimate the probability distributions of the output stochastic variables, which can only assume specific distributions, including Gaussian, Beta and Weibull in the subsequent study. The PDF and CDF curves of the system dynamic frequency deviation in the IEEE-39 Bus system are shown in Figs. 11-13.
\begin{figure}[htbp]
%\vspace{-0.1cm}
\centering
\subfigure[PDF]
{
    \begin{minipage}[b]{.4\linewidth}
        \flushleft
        \includegraphics[scale=0.13]{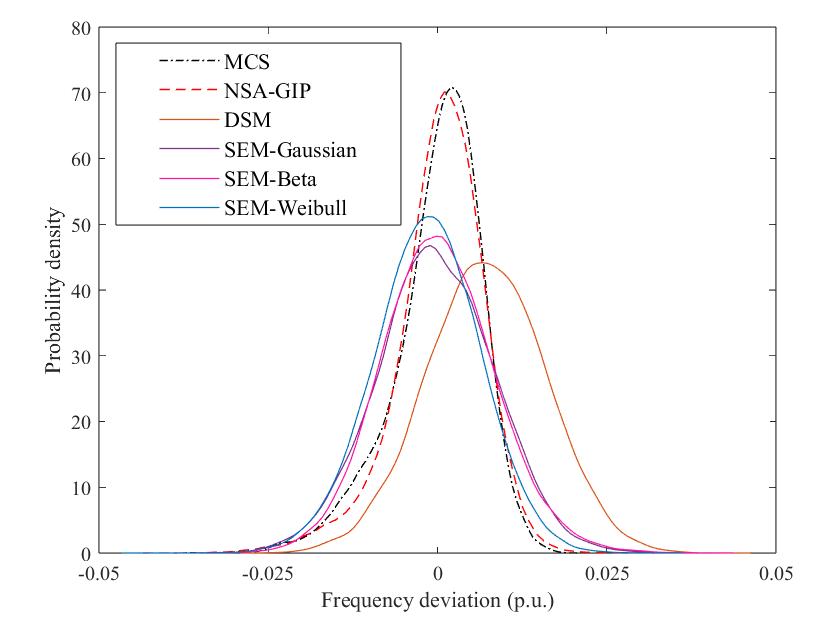}
    \end{minipage}
}
\subfigure[CDF]
{
    \begin{minipage}[b]{.4\linewidth}
        \flushleft
        \includegraphics[scale=0.13]{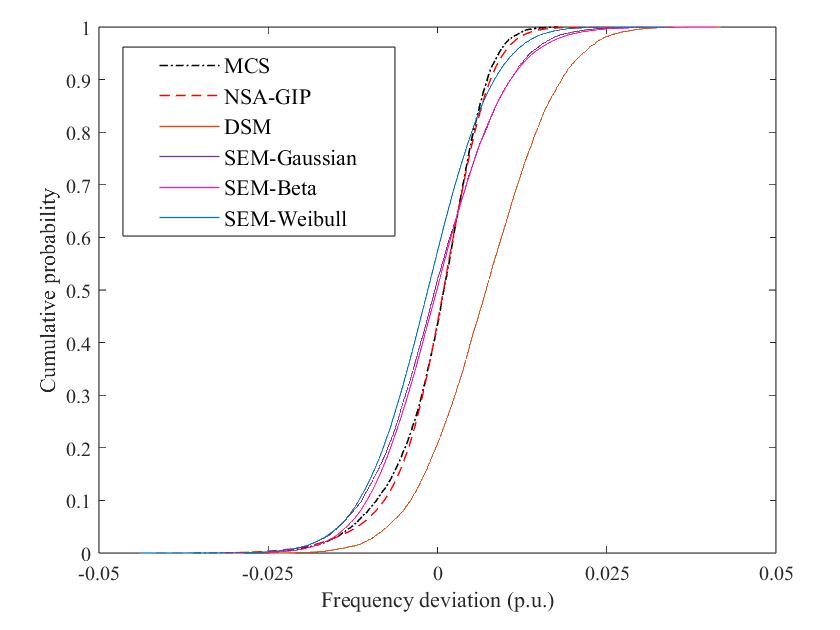}
    \end{minipage}
}
\vspace{-0.3cm}
\caption{Probability distribution of the system dynamic frequency deviation at 10s.}
\label{fig11}
\vspace{-0.4cm}
\end{figure}
\begin{figure}[htbp]
\vspace{-0.1cm}
\centering
\subfigure[PDF]
{
    \begin{minipage}[b]{.4\linewidth}
        \flushleft
        \includegraphics[scale=0.13]{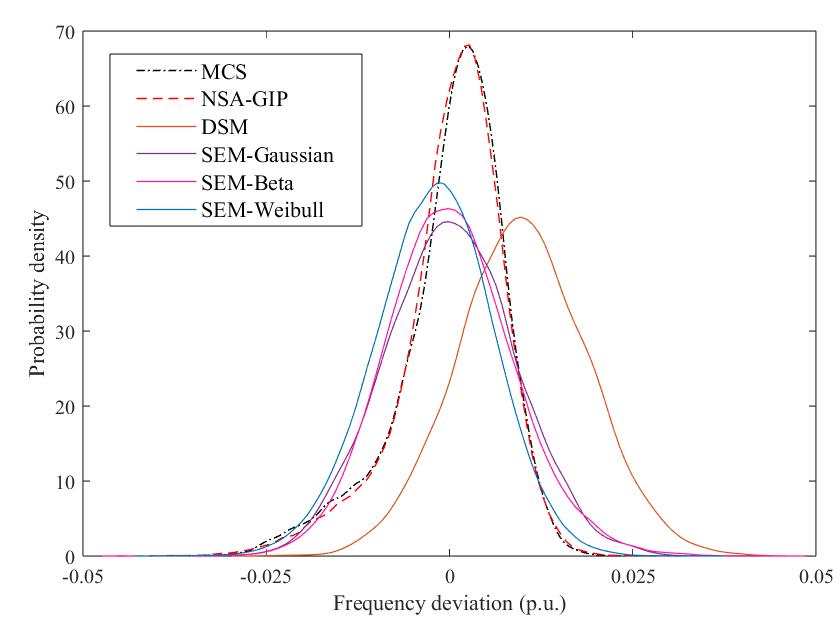}
    \end{minipage}
}
\subfigure[CDF]
{
    \begin{minipage}[b]{.4\linewidth}
        \flushleft
        \includegraphics[scale=0.13]{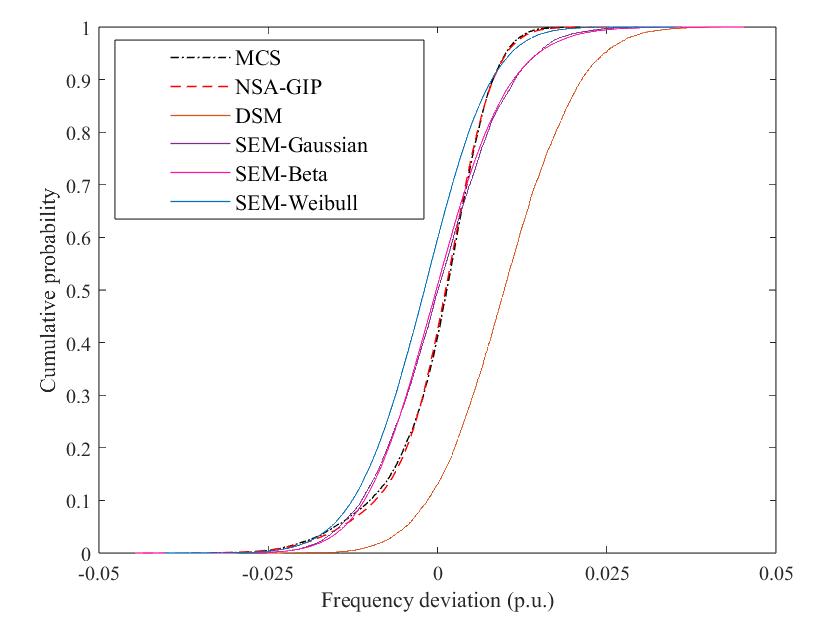}
    \end{minipage}
}
\vspace{-0.2cm}
\caption{Probability distribution of the system dynamic frequency deviation at 5s.}
\label{fig12}
\vspace{-0.5cm}
\end{figure}
\begin{figure}[htbp]
\vspace{-0.1cm}
\centering
\subfigure[PDF]
{
    \begin{minipage}[b]{.4\linewidth}
        \flushleft
        \includegraphics[scale=0.13]{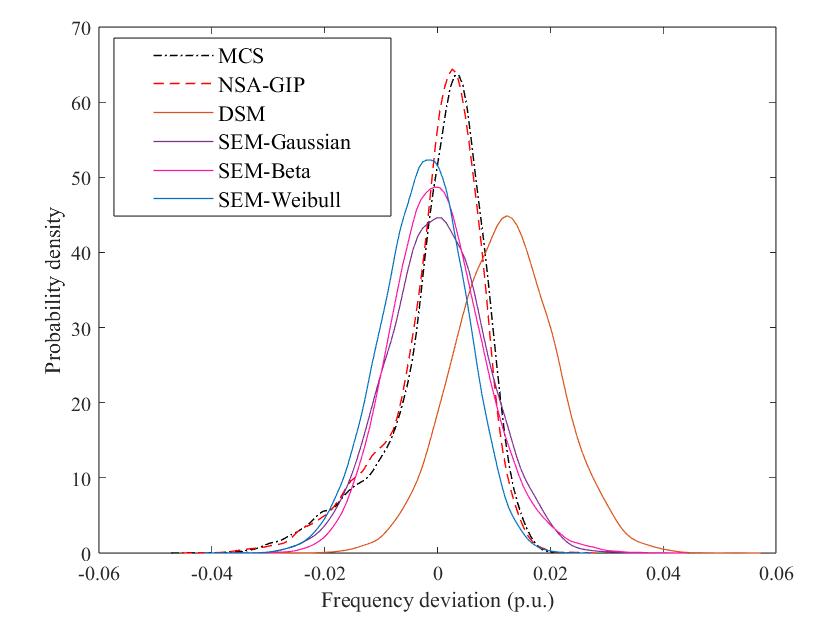}
    \end{minipage}
}
\subfigure[CDF]
{
    \begin{minipage}[b]{.4\linewidth}
        \flushleft
        \includegraphics[scale=0.13]{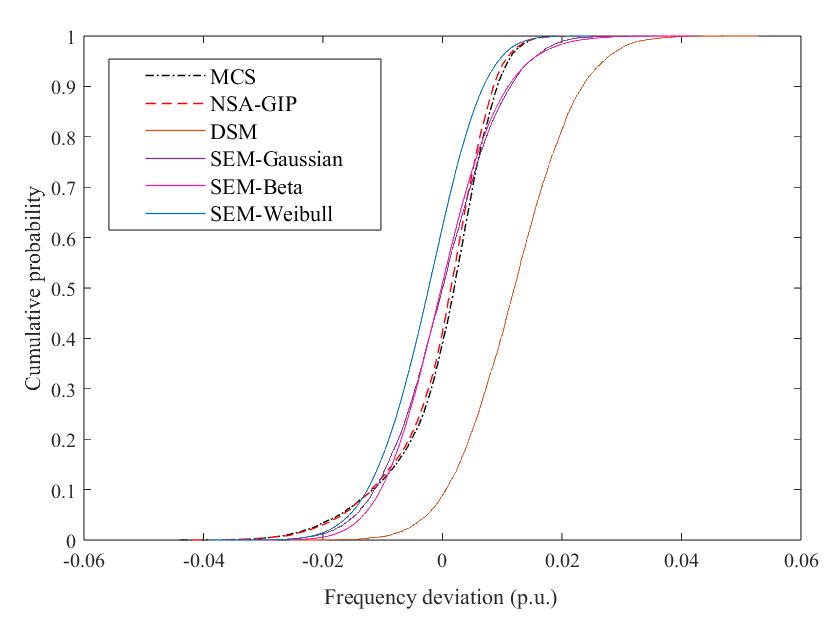}
    \end{minipage}
}
\vspace{-0.3cm}
\caption{Probability distribution of the system dynamic frequency deviation at 2.5s.}
\label{fig13}
\vspace{-0.1cm}
\end{figure}

Fig. 11 (a) and (b) show the probability density and cumulative probability curves of $\Delta f_t$ at 10s, respectively. Similarly, Fig. 12 and Fig. 13 show the probability density and cumulative probability curves of $\Delta f_t$ at 5s and 2.5s, respectively. It can be seen from these figures that the PDF and CDF of the system dynamic frequency deviation obtained by the NSA-GIP method is the closest to those obtained by MCS compared with the other two methods, regardless of any distribution assumption made by the other two methods.
\begin{table}[htbp]
\vspace{-0.3cm}
\caption{Standard Variance of Dynamic Frequency Deviation}
\label{tab2}
\centering
\vspace{-0.2cm}
\begin{tabular}{c c c c c c c}
\toprule[1.1pt]
$t (s)$ & MCS & NSA-GIP & DSM & \makecell[c]{SEM-\\Gussian} & \makecell[c]{SEM-\\Beta} & \makecell[c]{SEM-\\Weibull}\\
\midrule
0.5 & 0.0030 & 0.0031 & 0.0018 & 0.0022	& 0.0023 & 0.0020\\
2.5 & 0.0069 & 0.0067 & 0.0088 & 0.0089 & 0.0084 & 0.0075\\
5 & 0.0068 & 0.0066 & 0.0087 & 0.0088 & 0.0087 & 0.0079\\
7.5 & 0.0067 & 0.0066 & 0.0086 & 0.0087 & 0.0084 & 0.0078\\
10 & 0.0066 & 0.0065 & 0.0085 & 0.0086 & 0.0085 & 0.0078\\
15 & 0.0067 & 0.0066 & 0.0085 & 0.0086 & 0.0084 & 0.0077\\
\bottomrule[1.1pt]
\end{tabular}
\vspace{-0.1cm}
\end{table}

From the probability density and cumulative probability curves obtained by MCS, it can be seen that the probability distribution shape of the system dynamic frequency is significantly not close to any parametric distributions. However, the results obtained by DSM and SEM still have obvious Gaussian or other distribution characteristics, leading to significant approximation errors compared with MCS. In contrast, both the probability density and cumulative probability curves produced by the proposed NSA-GIP method in this paper is well consistent with MCS.

The standard variance of the system dynamic frequency deviation is shown in Table II. It can be found that the standard deviation obtained by DSM and SEM are significantly different from that of MCS, with the maximum relative errors more than 10\% regardless of the distributions assumed by SEM. The maximum relative error of NSA-GIP method is only about 4\%, which is significantly smaller than the results of DSM and SEM.

To further validate the accuracy of the proposed method, the PD curves at 5s and 10s are shown in Fig. 14, and the maximum PD values are shown in Table III. It can be seen that the PD values obtained by DSM deviate significantly from those of reference with the maximum PD value more than 25\%, and the maximum PD value of SEM is more than 9\% no matter what distribution assumed, which indicates that the probability characteristics of dynamic frequency cannot be accurately described by a specific parametric probability distribution. However, the maximum PD value of the NSA-GIP method is less than 4\%, indicating that it has significantly better accuracy performance than the existing methods.
\begin{figure}[htbp]
\vspace{-0.3cm}
\centering
\subfigure[5s]
{
    \begin{minipage}[b]{.4\linewidth}
        \flushleft
        \includegraphics[scale=0.13]{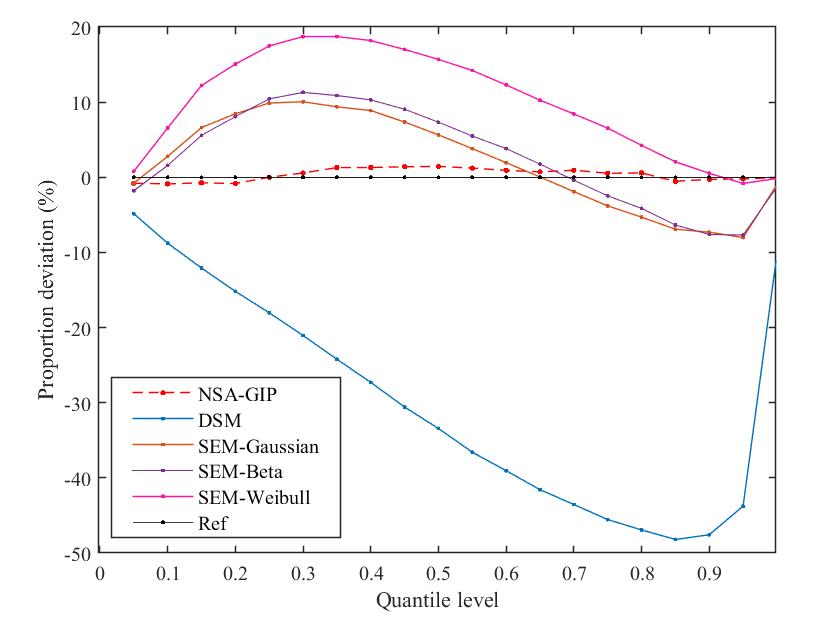}
    \end{minipage}
}
\subfigure[10s]
{
    \begin{minipage}[b]{.4\linewidth}
        \flushleft
        \includegraphics[scale=0.13]{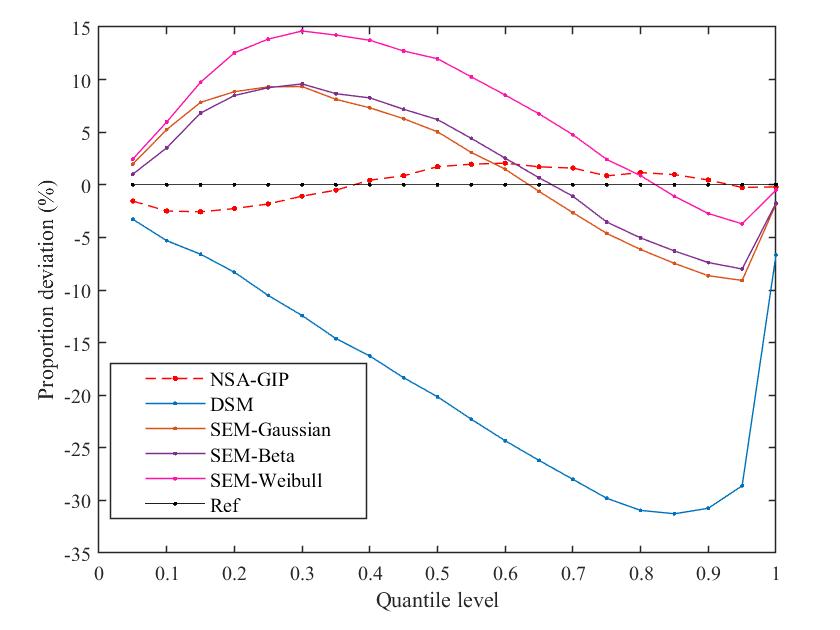}
    \end{minipage}
}
\vspace{-0.3cm}
\caption{Proportion deviation of the system dynamic frequency deviation at 5s and 10s.}
\label{fig14}
\vspace{-0.3cm}
\end{figure}

To further compare the differences of probability distributions obtained by different methods, Wasserstein distance is introduced as follow \cite{ref32}
\begin{equation}
\label{49}
W\left(P_1, P_2\right)=\inf _{\psi \sim \Pi\left(P_1, P_2\right)} E_{\left(x_1, x_2\right) \sim \psi}\left[\left\|x_1-x_2\right\|\right]
\end{equation}
where $W$ represents the Wasserstein distance between two different distributions $P_1$ and $P_2$, $\psi$ is the joint distribution of $P_1$ and $P_2$, $x_1$ and $x_2$ are two samples of $\psi$. $\left\|x_1-x_2\right\|$ represents the distance of $x_1$ and $x_2$.
\begin{table}[htbp]
\vspace{-0.3cm}
\caption{Maximum PD of Dynamic Frequency Deviation}
\label{tab3}
\centering
\vspace{-0.2cm}
\begin{tabular}{c c c c c c}
\toprule[1.1pt]
$t (s)$ & NSA-GIP & DSM & \makecell[c]{SEM-\\Gussian} & \makecell[c]{SEM-\\Beta} & \makecell[c]{SEM-\\Weibull}\\
\midrule
2.5 & 2.78\% & 31.95\% & 11.79\% & 12.31\% & 21.96\% \\
5 & 1.57\% & 38.17\% & 10.66\% & 11.35\% & 17.74\% \\
7.5 & 2.13\% & 38.23\% & 10.41\% & 9.86\% & 16.54\% \\
10 & 3.22\% & 32.37\% & 9.36\% & 9.54\% & 14.34\% \\
15 & 2.44\% & 27.62\% & 8.68\% & 9.05\% & 12.66\% \\
\bottomrule[1.1pt]
\end{tabular}
\vspace{-0.1cm}
\end{table}

Table IV shows the W-distance between the three methods and MCS. It can be seen that W-distance of NSA-GIP method is less than 0.0006, which is less than 25\% of SEM under three different distributions and 10\% of DSM. 

From the above analysis, it can be seen that traditional methods based on parametric probability assumptions cannot ensure the accuracy, as the actual probability distribution of wind power generation is extremely complicated. In contrast, the proposed method based on the generalized Ito process describes the uncertainty of the input power fluctuation and the output dynamic frequency in nonparametric form independent of any model assumption, which effectively improves the accuracy of stochastic analysis of system dynamic frequency.
\begin{table}[htbp]
\vspace{-0.3cm}
\caption{Wasserstein Distance of Dynamic Frequency Deviation}
\label{tab4}
\centering
\vspace{-0.2cm}
\begin{tabular}{c c c c c c}
\toprule[1.1pt]
$t (s)$ & NSA-GIP & DSM & \makecell[c]{SEM-\\Gussian} & \makecell[c]{SEM-\\Beta} & \makecell[c]{SEM-\\Weibull}\\
\midrule
2.5 & 0.0005 & 0.0078 & 0.0019 & 0.0022 & 0.0031\\
5 & 0.0004 & 0.0118 & 0.0019 & 0.0021 & 0.0025\\
7.5 & 0.0004 & 0.0103 & 0.0018 & 0.0019 & 0.0022\\
10 & 0.0005 & 0.0065 & 0.0018 & 0.0018 & 0.0019\\
15 & 0.0004 & 0.0053 & 0.0017 & 0.0016 & 0.0018\\
\bottomrule[1.1pt]
\end{tabular}
\vspace{-0.2cm}
\end{table}

\vspace{-0.3cm}
\subsection{Influence of Renewable Energy Penetration}
In order to analyze the influence of renewable energy penetration on the proposed method, different cases wind power penetration ranging from 20\% to 60\% are further studied with an increment of 10\%. The comprehensive estimation performance indexes of dynamic frequency uncertainty are shown in Table V. It can be seen that no matter how the renewable energy penetration changes, the maximum PD value can always be maintained below 4\% and the W-distance is within 0.0006, which indicates the high accuracy of the proposed method will not be influenced by in the increasing penetration of renewable energy generation.
\begin{table}[htbp]
\vspace{-0.3cm}
\caption{The Influence of Renewable Energy Penetration}
\label{tab5}
\centering
\vspace{-0.2cm}
\begin{tabular}{c c c}
\toprule[1.1pt]
Penetration (\%) & Maximum PD (\%) & W-distance\\
\midrule
20 & 2.57 & 0.0003\\
30 & 3.22 & 0.0005\\
40 & 3.12 & 0.0004\\
50 & 3.67 & 0.0005\\
60 & 3.42 & 0.0005\\
\bottomrule[1.1pt]
\end{tabular}
\vspace{-0.3cm}
\end{table}

\vspace{-0.3cm}
\subsection{Influence of GMM Component Number}
The influence of the Gaussian component number on the proposed method is shown in Table VI. It can be seen that the more Gaussian component number, the higher the solution accuracy, the longer the time required. However, when the number of Gaussian components is greater than a certain value, the accuracy does not differ much, so it is not necessary to select a very large number of Gaussian components. Furthermore, too many sub-Gaussian components may increase the risk of over-fitting and thus reduce the generalization performance. Therefore, the choice of the component number should consider many factors such as accuracy, efficiency and model generalization performance. In the study, the number of Gaussian components is set to 10.
\begin{table}[htbp]
\vspace{-0.3cm}
\caption{The Influence of Gaussian Component Number}
\label{tab6}
\centering
\vspace{-0.2cm}
\begin{tabular}{c c c c c c c}
\toprule[1.1pt]
GMM & 4 & 6 & 8 & 10 & 15 & 20\\
\midrule
$t(s)$ & 0.76 & 0.94 & 1.11 & 1.24 & 1.68 & 2.13\\
Maximum PD (\%) & 7.83 & 5.55 & 3.87 & 3.22 & 3.17 & 3.14\\
\bottomrule[1.1pt]
\end{tabular}
\vspace{-0.3cm}
\end{table}

\vspace{-0.3cm}
\subsection{Computational Efficiency Analysis}
The time required to complete the IEEE 39-Bus system with different methods is shown in Table VII. All simulations are conducted on a computer with an Intel Core i7-7700 CPU and 16 GB memory. MCS can ensure extremely high accuracy, but its calculation time is too long with more than 500s. DSM has the highest computational efficiency with only 0.51s. Compared with MCS, the SEM calculation time also has a very significant improvement, which only takes 1.28s. However, as aforementioned, the accuracy of DSM and SEM is limited due to the assumption of specific probability distribution model. In contrast, the proposed NSA-GIP method only needs significantly short calculation time 1.24s, while having excellent accuracy, which demonstrates high potential for real-time analysis applications in modern power systems with high penetration of renewable energy.
\begin{table}[htbp]
\vspace{-0.3cm}
\caption{Calculation Time of Different MethodS}
\label{tab7}
\centering
\vspace{-0.2cm}
\begin{threeparttable}
\begin{tabular}{c c c c c}
\toprule[1.1pt]
Method & MCS & NSA-GIP & DSM & SEM\\
\midrule
Calculation time (s) & 527.23 & 1.24 & 0.51 & 1.28\\
\bottomrule[1.1pt]
\end{tabular}
\begin{tablenotes}
            \item *The calculation time of SEM is the mean of time under various distribution assumptions.
        \end{tablenotes}
\end{threeparttable}
\vspace{-0.1cm}
\end{table}

The above results solidly show that the proposed NSA-GIP method ensures excellent accuracy and calculation efficiency. It can provide an effective support tool to ensure the frequency security of power systems with high penetration of renewables. The probability distribution of stochastic variables is described nonparametric based on the generalized It\^o process, which avoids the errors introduced by traditional methods when estimating the probability distribution through parametric assumptions or simple statistical moments, and greatly improves the calculation accuracy under the premise of high calculation efficiency.

\vspace{-0.1cm}
\section{Conclusion}\label{sec4}
The increasing penetration of renewable energy leads to more frequent frequency fluctuations in the power systems.  A novel nonparametric stochastic analysis method based on generalized It\^o process is developed to estimate the probability distribution of the system dynamic frequency under high-penetration renewable energy. The nonparametric probabilistic forecasting is firstly used to obtain the wind power probability distributions instead of parametric assumptions. A generalized It\^o process is proposed to describe any probability distribution, which can overcome the shortage that the classic Ito processes can only describe specific parametric distributions. A VSG-SFR stochastic model is constructed to consider the wind power frequency support. Based on the generalized It\^o process, the complex nonlinear SDE corresponding to the above model can be transformed into a linear combination of several linear SDEs, which can significantly reduce the solving difficulty. The accuracy and speed of the proposed method are compared with those of MCS and existing analytical methods, verifying its excellent computational efficiency. Further, the influences of the model parameters, renewable energy penetration and GMM component number on the proposed method is examined, which indicates that it is almost unaffected by the above three. In general, the proposed NSA-GIP method successfully analytically obtains the probability distribution of system dynamic frequency regardless of the distribution form of the input renewable generation, and has high application value in secure operation of power systems with high penetration of renewable energy.

\end{document}